\documentclass[prb]{revtex4-2}
\pdfoutput=1 

\usepackage{graphicx} 

\usepackage[utf8]{inputenc}
\usepackage{amsmath}
\usepackage{amssymb}
\usepackage{geometry}
\usepackage{physics}
\usepackage{xcolor}
\usepackage{caption}
\usepackage{subcaption}
\usepackage{url}

\newcommand{\voc}{\mathrm{V_{oc}}}
\newcommand{\Exx}{\ensuremath{\mathrm{E_{xx}}}}

\newcommand{\SubCap}[1]{\textbf{\textcolor{blue}{(#1)}}}

\newcommand{\BlockComment}[1]{}

\definecolor{orangered}{rgb}{1,0.4,0}
\definecolor{dred}{rgb}{0.83,0.15,0.15}
\definecolor{dgreen}{rgb}{0.1,0.5,0.1}
\definecolor{dyellow}{rgb}{0.45,0.45,0.0}
\definecolor{trq}{rgb}{0.05,0.45,0.45}
\definecolor{lblue}{rgb}{0.15,0.35,0.75}

\DeclareUnicodeCharacter{2215}{/}

\begin{document}

\title{Fermi Level and Light Driven Defect Generation in Silicon Solar Cells}

\author{Andrew \surname{Diggs}$^{\ddag}$}
\email{amdiggs@ucdavis.edu}

\author{Zitong \surname{Zhao}$^{\ddag}$}
\email{ztzhao@ucdavis.edu}
\affiliation{Physics Department, University of California\\
 Davis, CA 95616, USA}
\thanks{These two authors contributed equally to this work.}
\author{Adam \surname{Goga}}
\affiliation{Physics Department, University of California\\
 Davis, CA 95616, USA}
\author{Zachary \surname{Crawford}}
\affiliation{Physics Department, University of California\\
 Davis, CA 95616, USA}
\author{Gergely T. \surname{Zim\'anyi}}
\affiliation{Physics Department, University of California\\
 Davis, CA 95616, USA}

\date{January 11, 2025}

\begin{abstract}
Hydrogenated amorphous silicon (a-Si:H) has had a long standing role as a passivating dielectric for c-Si, often utilized in the early development of ICs and more recently for Si solar cells. Although it has been studied for more than 60 years, several questions about the material properties remain open, including light-induced degradation and the Fermi level dependence on the mobility of hydrogen. Here we study the origin of these phenomenon using electronic structure based calculations. First, we use density functional theory (DFT) and the nudged elastic band (NEB) method to examine defect generation via Si-H bond breaking in p-type, intrinsic, and n-type a-Si:H. We find that the energy barrier controlling this defect generation, shows the same asymmetric reduction of $\sim 0.3$ eV for p-type and $\sim 0.1$ eV for n-type, observed in experimental studies. We then develop a model based on the local Coulomb interactions at the transition state, which provides compelling evidence that the asymmetry results from the emergence of a high energy donor state created by the interstitial H. Finally, we repeat our Si-H bond breaking analysis, combining NEB with constrained density functional perturbation theory (c-DFPT) to simulate defect generation dynamics in illuminated a-Si:H. Here we find that e-p pair results in a combined effect that reduces the barrier by $\sim 0.4$, in close agreement with experimental observations.
\end{abstract}
\maketitle
\section{Introduction}

Silicon heterojunction (Si HJ) solar cells are part of the promising ``next generation'' Si solar cells, which utilize a passivating contact cell designs. Si HJ currently hold the efficiency record for single crystal non-concentrator Si solar cell, at 27.3\% \cite{NREL_Efficiency_Chart}, largely due to their unmatched open circuit voltage ($\voc$). A key driver for their performance is the excellent electronic passivation of the c-Si absorber by a thin layer of intrinsic hydrogenated amorphous silicon below the doped contact layer. In spite of their great promise, fielded modules have shown an initial degradation of $\sim 0.7\%$ per yr \cite{NRELdegradation}, much higher than the usual rate of 0.2\% per yr \cite{GAO2022}. 
The excess 0.5\% per yr rate was attributed to the degradation of the open circuit voltage $\voc$ \cite{NRELdegradation,GAO2022}, which has been linked to the loss of passivating hydrogen \cite{SolDegH}. Therefore, understanding and subsequently minimizing the performance loss of Si HJ solar cells is key to accelerating their market acceptance.

Amorphous silicon has had a long standing role as a passivating dielectric, often utilized in the early development of ICs.
Although it has been studied for more than 60 years, several questions about the material properties remain open, such as understanding the classes of structural and electronic defects, their formation and statistical distributions, and the long-term structural dynamics. A comprehensive review of these issues was compiled by N. M. Johnson in 1991 \cite{JOHNSON1991}, and more recent reviews of these issues specifically relating to solar cells are those by S. DeWolf 2012 \cite{DeWolf2012Review} and C. Wronski 2014. \cite{Wronski2014Review}   

One of the more intriguing unresolved issues is light induced degradation in intrinsic amorphous silicon, commonly referred to as the Stabler-Wronski effect (SWE).
In 1977 Staebler and Wronski found that after prolonged light exposure, intrinsic a-Si:H underwent a losses of the dark conductivity. This degradation was fully reversible through dark annealing. It was suggested that the decrease in the dark conductivity is due to a decrease in the thermal free carrier density which is strongly linked to the Fermi level. The drop in the conductivity/Fermi level was attributed to an increase in midgap states \cite{WRONSKI1984}.
This phenomenon has been the focus of many studies over the last 50 years \cite{Olibet2006_LID,StutzmannLID1985,Elliot1979,Crandall1991,Melskens-Smets-2104, biswas,DBrecombination}; however, to this date there remains no agreed upon theory for the light induced degradation in a-Si:H. 
Many speculate that the primary driver of light induced degradation (LID) is generation of Si DB through the loss of hydrogen \cite{Santos-Johnson-Street-1991,Mahtani2013LID}.

This intuition has driven many studies to focus on the kinetics of hydrogen in a-Si:H. 
Specifically, large efforts have been made to understand the influence of illumination and forward bias injection, Fermi level, and electric fields on the mobility of hydrogen in a-Si:H \cite{BEYER1991,BEYER1996,Street-Kakalios-1987,Santos-Johnson-Street-1991,Santos-Johnson-Street-1992,Santos-Johnson-Street-1993,Kakalios-Street-Dispersive, biswas,ww,Herring2001}.

\begin{figure}[h]
    \centering
    \begin{subfigure}[t]{0.35\textwidth}
    \caption{}
    \includegraphics[width = \textwidth]{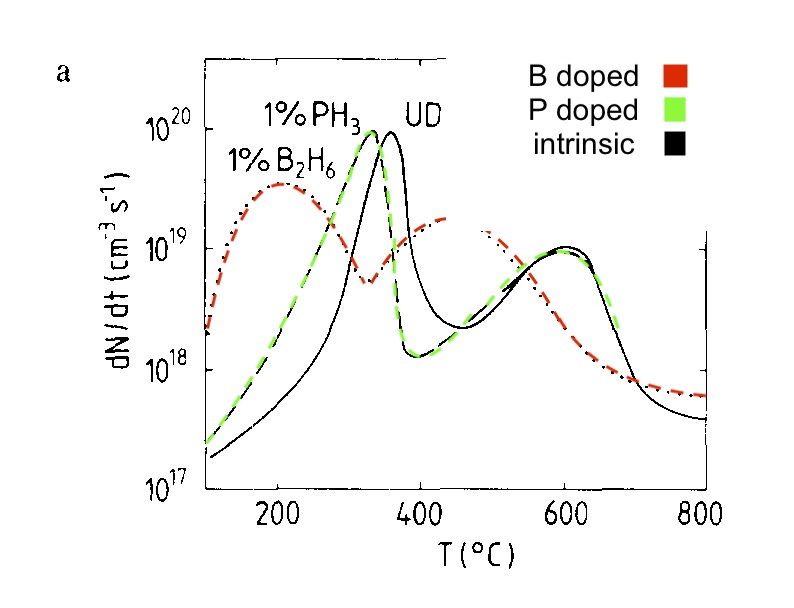}
    \label{fig:beyer}
    \end{subfigure}
    ~
    \begin{subfigure}[t]{0.60\textwidth}
    \caption{}
    \vspace{5pt}
    \includegraphics[width =\textwidth]{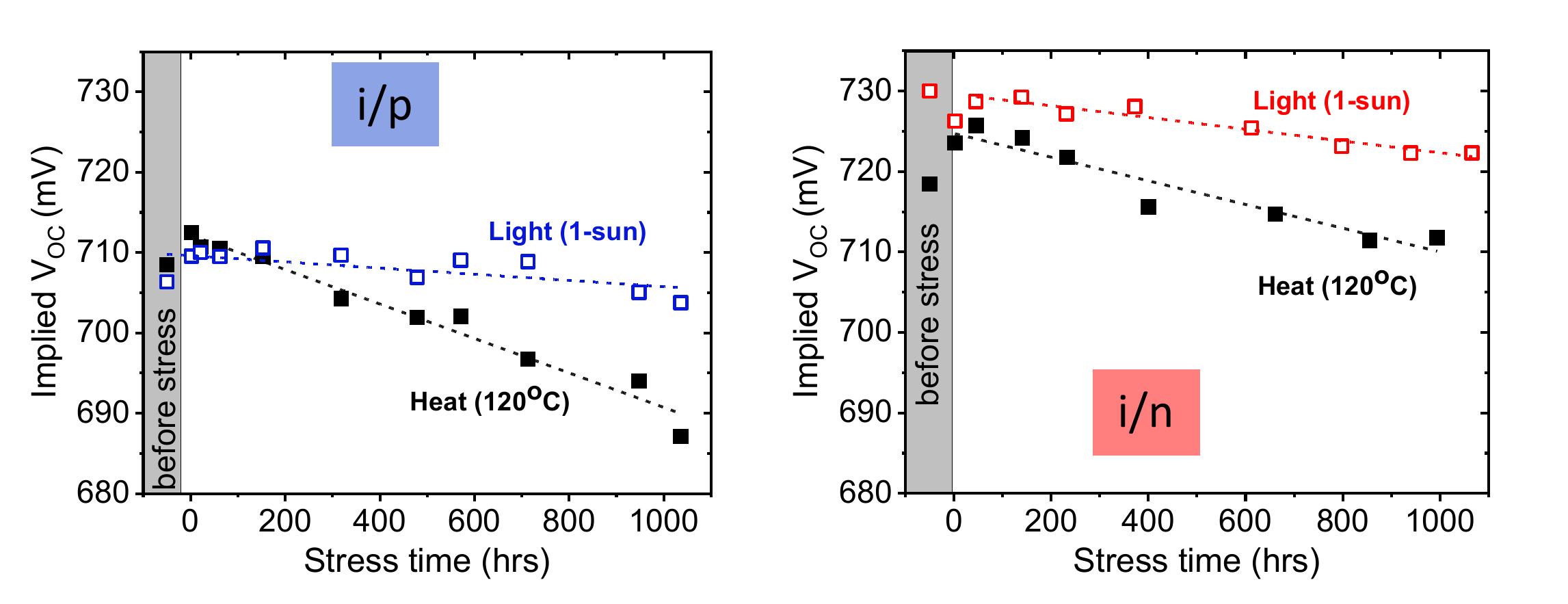}
    \label{fig:das}
    \end{subfigure}
    \vspace{-15pt}
    \caption{Experimental results showing the difference in hydrogen kinetics in a-Si:H and Si HJ. \textbf{a)} The effusion and diffusion of H in a-Si:H shows asymmetrical increases for p-type and n-type samples relative to intrinsic. Reproduced from  W. Beyer \textit{J. Non-Cryst. Solids} (1989). \textbf{b)} The thermal degradation in Si HJ stacks is noticeably different for n/i/n and n/i/p stacks. Courtesy of Ujjwal Das and Margaret Zeile IEC-University of Delaware.}
    \label{fig:high-low}
\end{figure}

From these studies additional mysterious and unresolved phenomena emerged, involving the charge state of mobile hydrogen and the influence of the Fermi level on the mobility of hydrogen. Figure \ref{fig:beyer} shows a well known result from the works of W. Beyer in 1991, who found that the effusion, and thus diffusion, of hydrogen is asymmetrically increased for doped a-Si:H, with a larger increase for p-type and a smaller increase for n-type \cite{BEYER1991}. Following these results, additional studies directly measured the diffusion of H in doped a-Si:H. These studies determined the energy barrier controlling hydrogen diffusion, and found a remarkable $\sim 0.3$ eV decrease in p-type a-Si:H and a $\sim 0.1$ eV decrease in n-type a-Si:H, relative to intrinsic a-Si:H. Additionally, compensated a-Si:H, B + P doped, showed almost no change in the mobility of H compared to intrinsic a-Si:H. The similar behaviors in intrinsic and compensated samples indicated that the change in hydrogen mobility was not tied to specific interactions with the dopent atoms but a direct result of the Fermi level \cite{Street-Kakalios-1987, BEYER1991, BEYER1996, BEYER2003, Nickel-Beckers2002}. The connection between hydrogen mobility and Fermi level sparked numerous efforts exploring this topic, many of which focused on the charge state of hydrogen as a possible explanation for this phenomenon \cite{Herring2001, Nickel-Beckers2002,zhu1990negU, vdw1994charge}. Recently this topic has reemerged after a similar phenomenon was observed regarding the degradation of the $\voc$ in Si HJ solar cells. The results of one of these studies is shown in Fig.~\ref{fig:das}, where a significant difference was found between the thermal degradation in n-type (lower) and p-type (higher) symmetrically passivated Si HJ stacks \cite{UDel_Deg}.

All of these issues play a crucial role in the passivation and degradation of the crystalline/amorphous interface of Si HJs \cite{Bertoni2019,Holovsky2020,soldeg}.
The primary form of degradation seen in Si HJ cells has been attributed to an increase in surface recombination at the c-Si/a-Si:H interface. Here, dangling bonds (DBs) are thought to be the primary recombination centers \cite{DBrecombination}. In a previous SolDeg study we developed a comprehensive model for the thermally driven degradation resulting from the loss of passivating hydrogen resulting in an increase of the DB density at the c-Si/a-Si:H interface \cite{SolDegH}. As a natural continuation of that work we have carried out a comprehensive first principals computational investigation of the influence of the Fermi level and photo-excited carriers on the mobility of hydrogen and defect generation in Si HJ. Specifically, we looked at defect generation via Si-H bond breaking in p-type, intrinsic, and n-type crystalline and amorphous Si. We found that near an available Si DB defect state, the lowest energy interstitial site for H in all types is the Si-Si bond center (BC), and that this bond center hydrogen (BCH) creates a high energy state near the conduction band tail which acts an electron donor. Additionally, we determined the energy barriers controlling this defect generation, where we found the same asymmetric reduction of $\sim 0.3$ eV for p-type and $\sim 0.1$ eV for n-type, compared to intrinsic Si. Furthermore, we found that the changes in these barriers can be well explained by the local Coulomb interactions, and that the asymmetry arises from the emergence of the BCH donor state. Finally, we repeat our Si-H bond breaking analysis, using constrained density functional perturbation theory (c-DFPT) to simulate electron-hole pairs generated by illumination. Here we find that the electronic configuration is such that the barrier gets the combined effect of both the p-type and n-type cells, resulting in a total reduction of $\sim 0.4$, which is in agreement with previously reported experiment results \cite{Santos-Johnson-Street-1991}.

\section{Defect Generation Dynamics in Crystalline and Amorphous Silicon}

We extended our SolDeg platform to study the local and global effects of the Fermi level on defect generation in silicon heterojunction solar cells by combining MD, DFT, full device simulations, and kinetic modeling. 
To carry out this large-scale investigation, simultaneous studies using DFT were conducted, one using a plain wave basis and the other using a Gaussian basis. Gaussian basis calculations were performed in the CP2K suite, using a Perdew-Burke-Ernzerhof (PBE) \cite{PBE} exchange correlation functional ($\mathrm{E_{xx}}$) and a Goedecker-Teter-Hutter (GTH) psuedopotential (PP). These calculations were performed to extend our capabilities to the $\sim 400$ atom c-Si/a-Si:H stacks created in our previous study \cite{SolDegH}, and to see if/how the contrasting localizing tendencies of the two basis sets might affect the results.\footnote{Plane waves are highly delocalized, and Gaussians are highly localized} 
Unless stated otherwise, the results presented here were performed using a plane wave basis in the Quantum Espresso (QE) suite \cite{QE1,QE2} using a PBEsol $\Exx$  \cite{PBESOL} and a projector augmented wave (PAW) PP \cite{PAW,PP_LIB}. Additional plane wave calculations were performed using a PBE0 hybrid functional with 0.25 exact exchange \cite{PBE0}, as well as Hubbard corrected density functional theory, with and without inter-site corrections (DFT+U+V)/(DFT+U) using a PBE $\Exx$ and a norm conserving PP \cite{lsda_U,Campo_2010dft_u_v,DFPT_U}.

\subsection{Defect Formation Kinetics in Doped c-Si}
\begin{figure}
    \centering
    \includegraphics[width=0.5\linewidth]{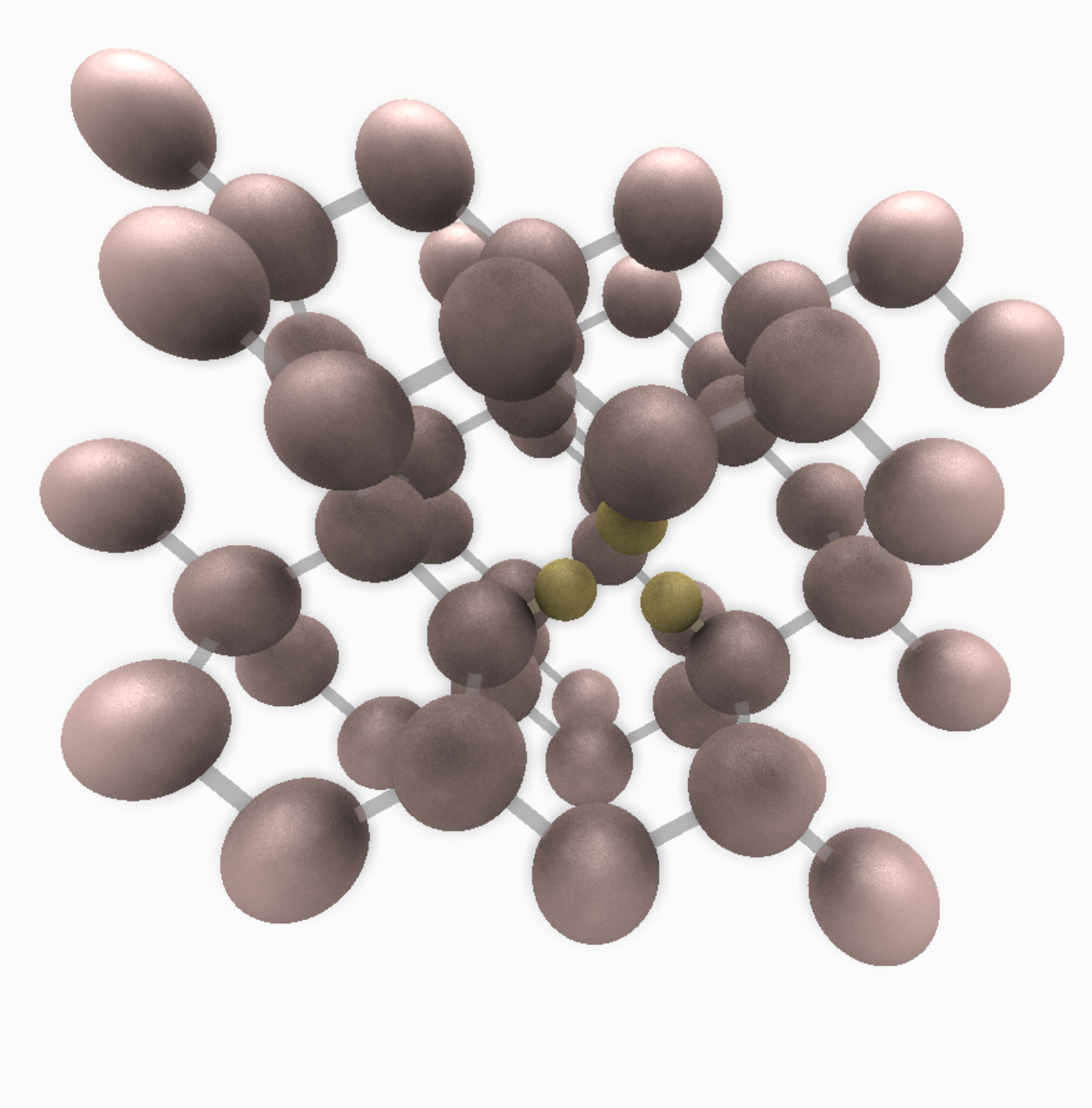}
    \caption{The initial structure used in this part of our work was a 63 Si + 4 H containing a fully passivated Si vacancy.}
    \label{fig:csi_im1}
\end{figure}

We started this study by exploring the effect of the Fermi level on defect formation in c-Si. To do this, we started with a $2\times2\times2 \;(5.43 \text{\r{A}})^3$ c-Si super cell containing 64 Si. One Si was removed, leaving four DB which were then passivated with hydrogen, resulting in a super cell containing 63 Si and 4 H, shown in Fig \ref{fig:csi_im1}. A similar structure using a $3\times3\times3 \;(5.43 \text{\r{A}})^3$ c-Si cell containing 215 Si and 4 H were used used for additional calculations to check for finite size effects.

To set the Fermi level in a simulation we used two methods: First, we added dopent atoms; specifically, B, nothing, P, or B+P, to create p-type, intrinsic, n-type, and compensated cells. Second, we varied the total super cell charge by subtracting or adding one electron from the neutral super cell, resulting in three cells with total cell charge $Q = +q_e, 0, -q_e$, representing p-type, intrinsic, and n-type c-Si. For clarity, in our notation the charge of an electron is given as $-q_e$. Thus, adding one electron results in a total cell charge $Q=-q_e$, and subtracting one electron, adding a hole, results in a cell with total cell charge $Q=+q_e$. Early in this study we found close correspondence between both methods. For the sake of simplicity the results presented here will be for the charged cell calculations only, unless stated otherwise. For specific results comparing both methods see Supplementary Materials.

To obtain our final defect-free initial cells, we performed a variable cell relaxation using BFGS optimization to an energy tolerance of $10^{-4} Ry$ and a force tolerance of $10^{-3} Ry$. Self consistent field (scf) calculations were performed using a kinetic energy cutoff of 50.0 Ry for wavefunctions and 400.0 Ry for charge density, with a convergence threshold of $10^{-8}$. K point sampling was done using a $4\times4\times4$ Monkhorst-Pack grid centered at $(0,0,0)$ for the 67 atom super cell, and a $2\times2\times2$ Monkhorst-Pack grid centered at $(0,0,0)$ for the 219 atom super cell. All parameters were established through convergence to an error tolerance of $< 0.1$ meV per atom.  

\begin{figure}
     \centering
    \begin{subfigure}[t]{0.30\textwidth}
    \caption{}
    \vspace{10pt}
    \includegraphics[width = \textwidth]{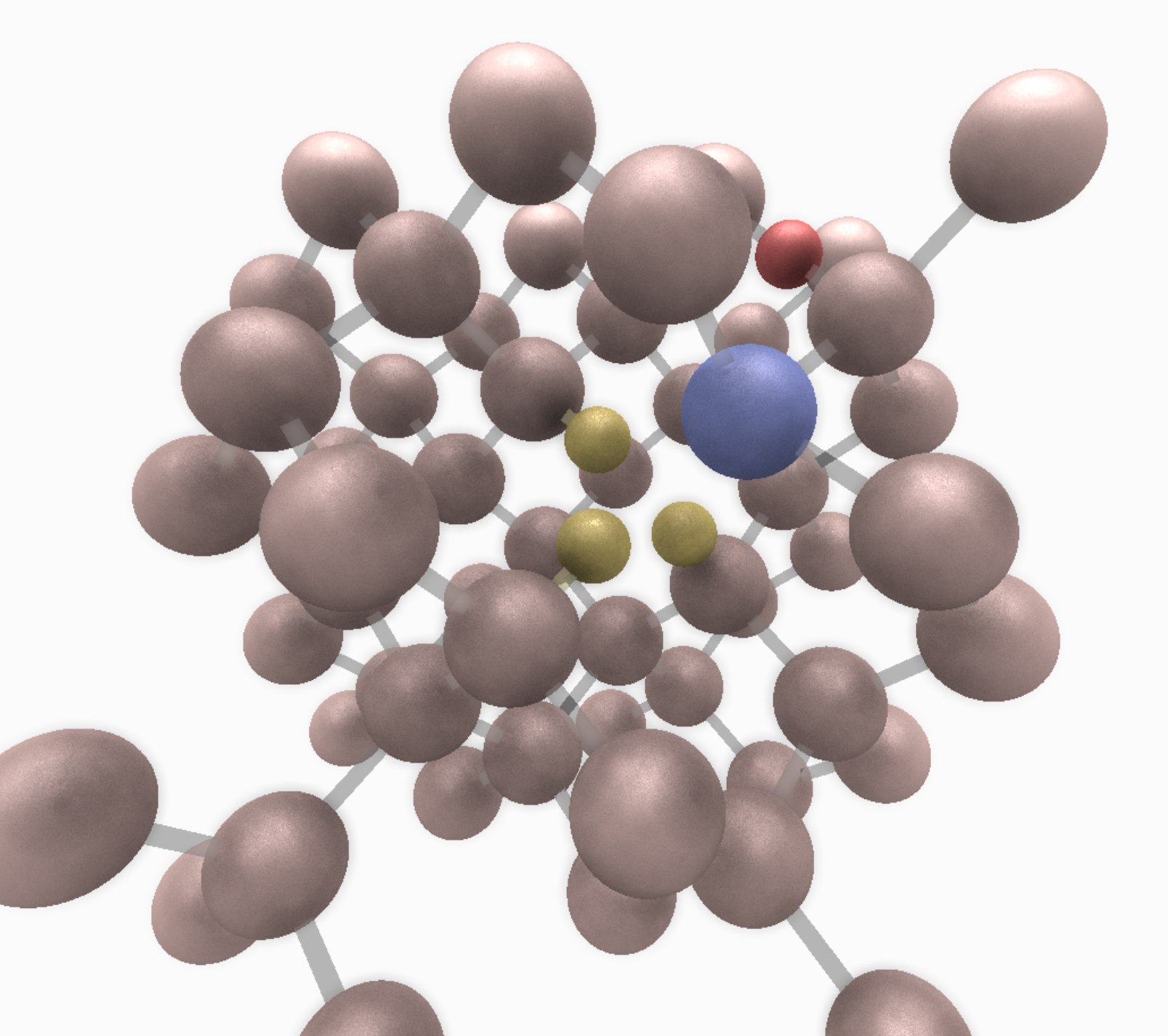}
    \label{fig:BCH}
    \end{subfigure}
    ~
    \begin{subfigure}[t]{0.30\textwidth}
    \caption{}
    \vspace{10pt}
    \includegraphics[width =\textwidth]{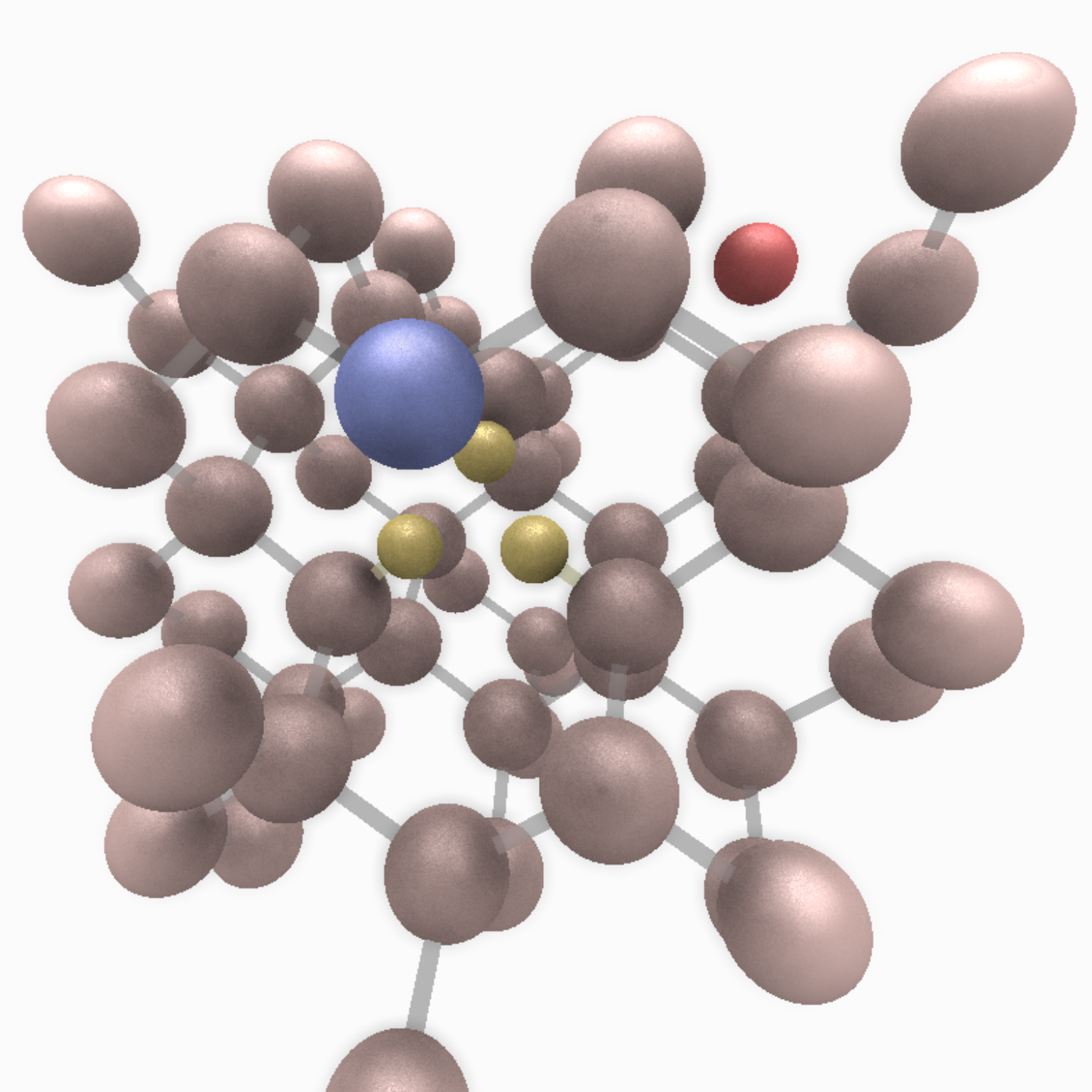}
    \label{fig:TET}
    \end{subfigure}
    ~
    \begin{subfigure}[t]{0.30\textwidth}
    \caption{}
    \vspace{10pt}
    \includegraphics[width =\textwidth]{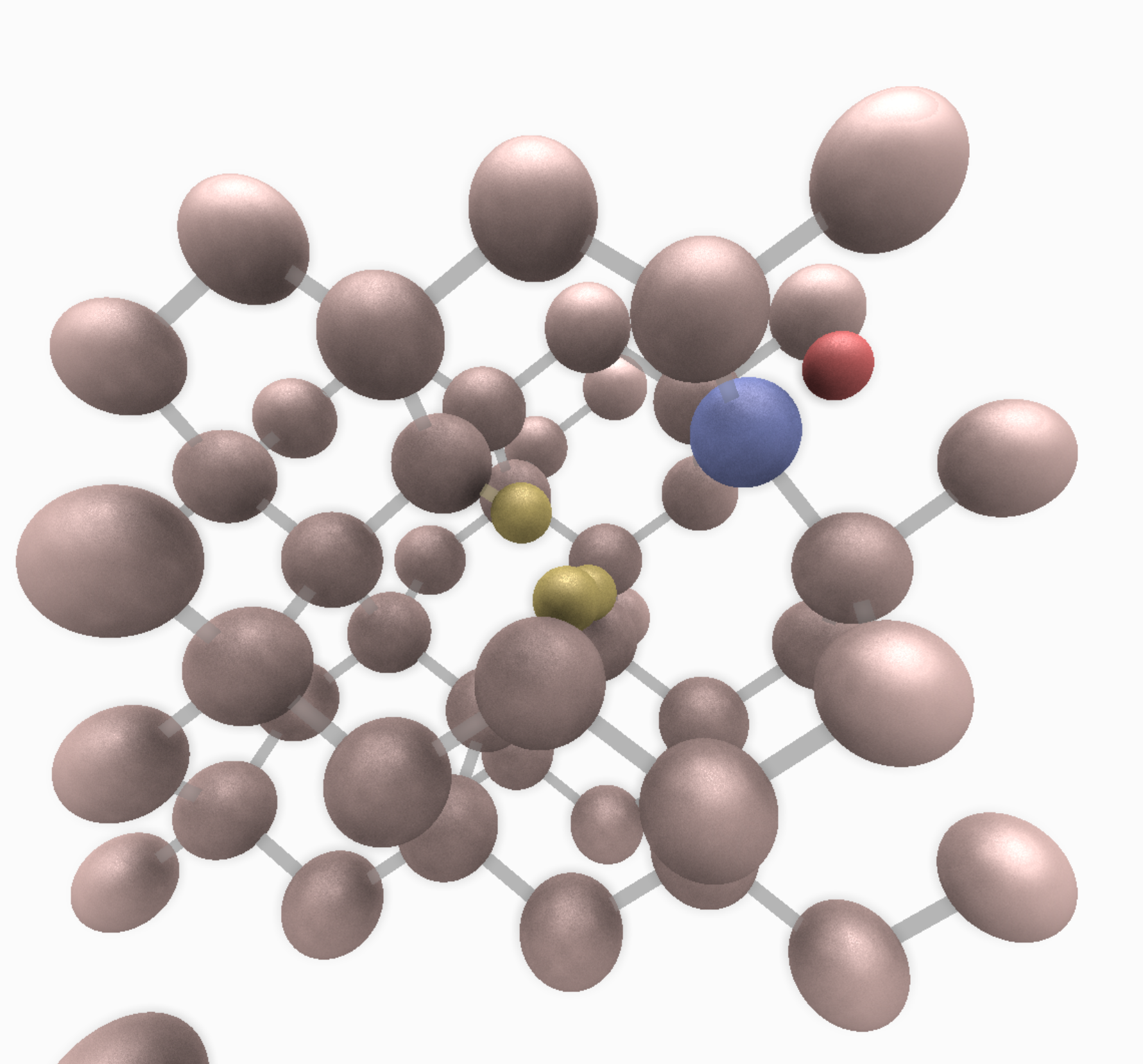}
    \label{fig:AB}
    \end{subfigure}
    \caption{Three interstitial hydrogen configurations, \SubCap{a} Si-Si BC, \SubCap{b} tetrahedral center, and \SubCap{c} the anti-bonding site, were tested for all three super cell charges $+q_e, 0, -q_e$, to establish the Fermi level dependent defect configuration. In agreement with previous work, for p-type and intrinsic, $+q_e$ and $0$, the Si-Si BC was the lowest energy defect configuration. In defect-free c-Si the tetrahedral center was lowest energy configuration for n-type, $-q_e$ \protect{\cite{Herring2001,JechGrasser2019}}. However, near the vicinity of a Si DB, the lowest energy configuration in the n-type cell shifted to a Si-Si BC.}
    \label{fig:defects}
\end{figure}

In our previous work we created defects by displacing a hydrogen from a Si-H bond to a next nearest neighbor Si-Si BC, which had been reported to be the most stable interstitial site for hydrogen in intrinsic Si \cite{Herring2001,VDW-Street-1994}, and was confirmed by our simulations. However, it has been reported that the lowest energy configuration for interstitial H is dependent on the Fermi level \cite{Herring2001}.
Based on an informed decision \cite{SolDegH,JechGrasser2019,Pantelides-2006,Herring2001} we explored the three most likely interstitial configurations: the Si-Si bond center, the tetrahedral center, and anti-bonding site (AB), shown in Fig.~\ref{fig:defects}. All three interstitial sites were tested for each Fermi level. Additionally, we carried out the same procedure in defect-free, 64 Si + H, intrinsic and n-type cells.

In all cells we found that the AB site was the lowest in energy. However, consistent with previous work \cite{Jech_Thesis}, we found that this configuration was only meta-stable and did not result in any recombination active defects.
In the p-type and intrinsic cells, the Si-Si BC was the lowest energy interstitial defect configuration for hydrogen, which agreed with previous studies \cite{Herring2001,VDW-Street-1994,SolDegH}. Previous works reported that for n-type Si the tetrahedral position was the lowest energy interstitial site for H \cite{Herring2001}. In the n-type cells, we found that the tetrahedral center was the lowest energy interstitial configuration for the defect-free cell; however, with a Si DB present, the lowest energy configuration shifted to the Si-Si BC. We will henceforth refer to these Si-H-Si bond centered hydrogen as (BCH) complexes.

\subsection{Identifying Recombination Active Defects}

\begin{figure}
    \centering
    \begin{subfigure}[t]{0.45\textwidth}
    \caption{}
    \vspace{0pt}
    \includegraphics[width = \textwidth]{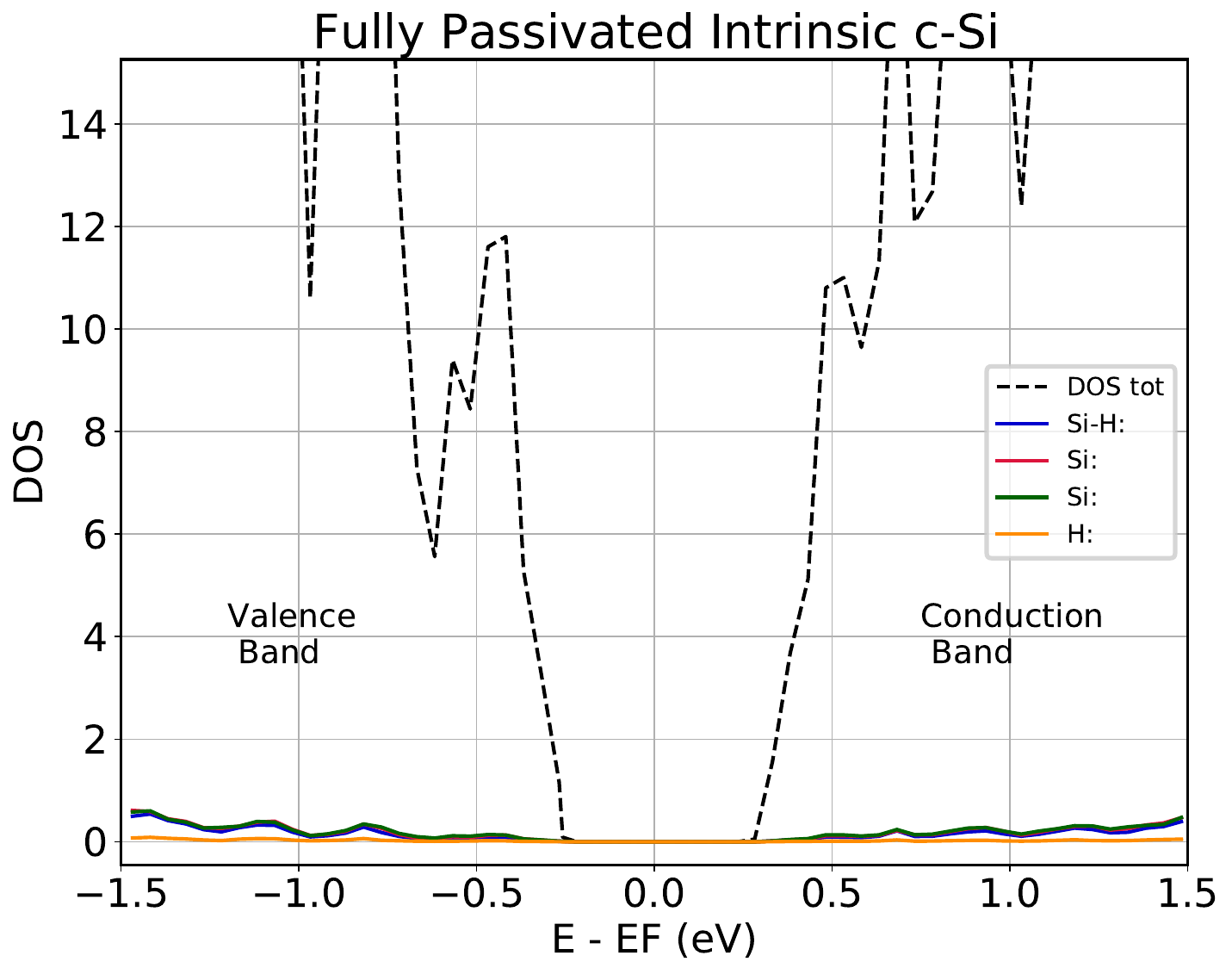}
    \label{fig:Q0_NoDB_PDOS}
    \end{subfigure}
    ~
    \begin{subfigure}[t]{0.45\textwidth}
    \caption{}
    \vspace{0pt}
    \includegraphics[width =\textwidth]{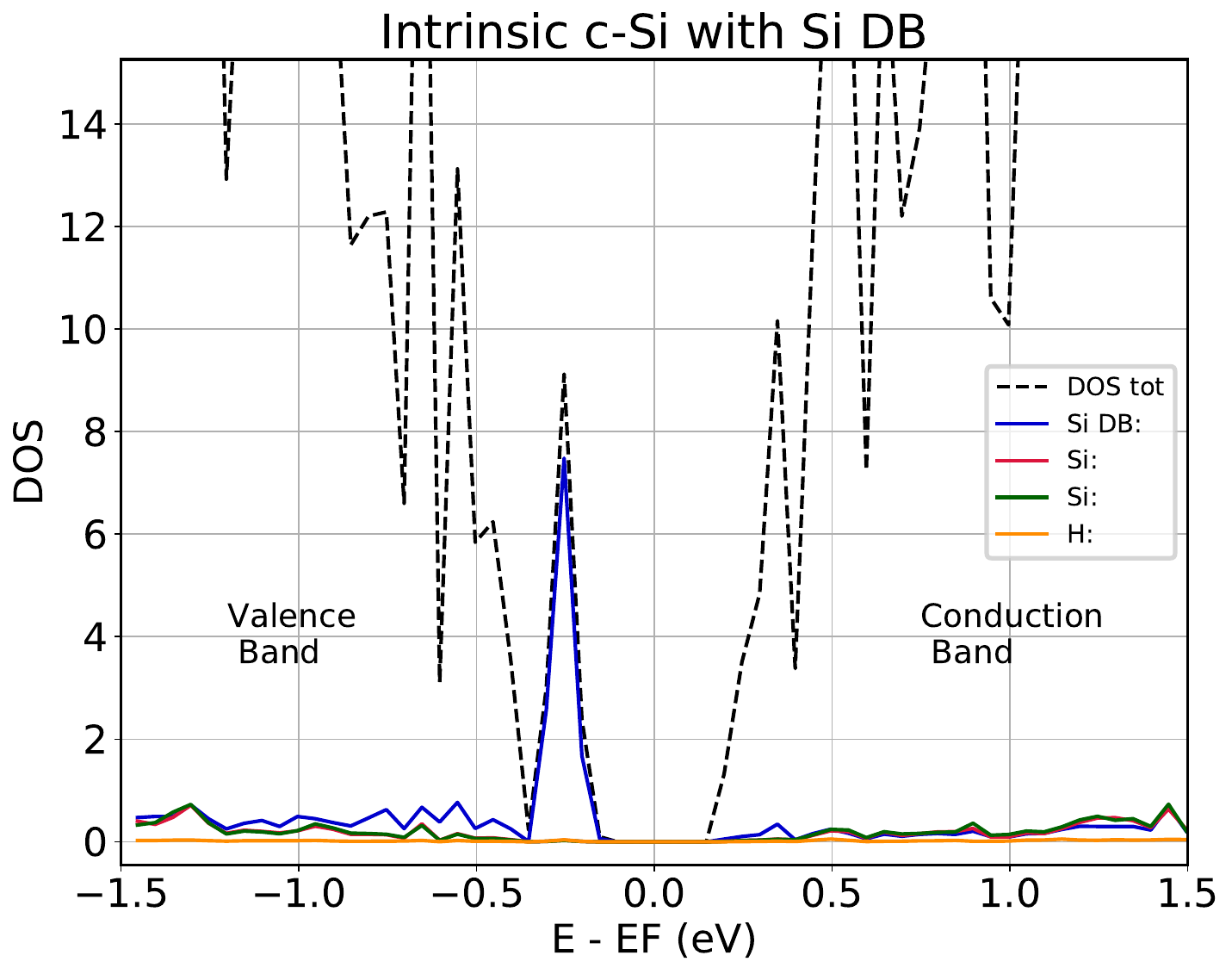}
    \label{fig:Q0_DB_PDOS}
    \end{subfigure}
    \vspace{-10pt}
    \caption{The projected density of states for \SubCap{a} the fully passivated c-Si cell which shows an uninterrupted bandgap indicating no electronic defects. The delocalized states are made from small contributions by many atoms, thus the density of a Si in these states(blue, red, and green lines) is much smaller than the total density of states (dashed black line). \SubCap{b} After the Si-H bond breaks a defect state appears in the bandgap. The projections show that this defect state is almost entirely localized on the Si DB(blue line). The states below the Fermi level are the occupied states. Unexpectedly for intrinsic Si, the Si DB is occupied by two electrons.}
    \label{fig:befor-after-PDOS}
\end{figure}

After finding the lowest energy configuration for interstitial hydrogen in our simulation cells we proceeded to determine the electronic properties of these configurations. 
To do this we computed the density of states projected onto the atomic orbital basis (PDOS), as implemented in QE. By projecting the DOS onto the atomic orbital basis we can directly discern the contribution of each individual atom to a state. A fully delocalized state contains contribution from many atoms, thus the projection should be small, $\sim \frac{1}{N}$. In contrast a localized state will have large contribution from only a few atoms. This is the same principal used for determining defects with the IPR. 
This technique allowed us to track the change in defect states between our initially fully passivated cells and our final cells containing one Si DB and one interstitial H, as well as determine the contributions to defect states by individual atoms. Figure \ref{fig:befor-after-PDOS} shows that in the fully passivated c-Si cell the bandgap is uninterrupted, thus no electronic defects are present. In the final configuration a midgap defect state appears, and it is almost entirely localized on the Si DB. 

One entirely unexpected result came out of our defect analysis. In Fig.~\ref{fig:befor-after-PDOS} the state of the Si DB is the state for one electron with degeneracy of two. For a spin symmetric calculation, only the lower half of the Si DB state would be occupied if the Si DB was occupied by one electron. This would result in a computed Fermi level, halfway between HOMO and LUMO, in the middle of the DB states. Figure \ref{fig:befor-after-PDOS} shows that in the final configuration the Fermi level is between the DB states and the conduction band, indicating that the DB is in fact occupied by two electrons.

\subsection{The Si-H-Si Donor State}

The Si DB is well known to be a deep amphoteric defect, meaning it can have three different occupations/charge states: empty, $1e^-,$ and $2e^-$ with corresponding charge $ q_{DB} =+q_e, 0,$ and $-q_e$. 
The energy of $q_{DB} =+q_e, 0$ charge states ($E_{DB_{+/0}}$) is near midgap, showing a negative charge transition level, $E_{DB_{-}} = \epsilon_{DB_{+/0}} + U$, with a reported correlation energy $U \approx 0.2$ eV \cite{Olibet2007,DeWolfKondo2007,vdw1994charge}. 
For a defect of this nature one would expect the ground state occupation to follow\footnote{Fermi-Dirac statistics describe only single electron occupations, and are not valid for double occupations with on-site energy. However, for a ground state configuration $E_F$ is specifically the Fermi energy and given by the HOMO and LUMO \cite{Shockley-Sah1958}.}

\begin{equation}
q_{DB} = 
\begin{cases}
    +q_e & E_F < \epsilon_{DB_{+/0}} \\
      0 & \epsilon_{DB_{+/0}} < E_F < \epsilon_{DB_{-}} \\
      -q_e & E_F > \epsilon_{DB_{-}}
\end{cases}  
\label{eqn:qDB}
\end{equation}

After observing the unexpected double occupation of the Si DB in the our defect structures, we proceeded to clarify these results prior to proceeding with our investigation into defect generation dynamics. 

\begin{figure}[h]
     \centering
    \begin{subfigure}[t]{0.30\textwidth}
    \caption{}
    \vspace{0pt}
    \includegraphics[width = \textwidth]{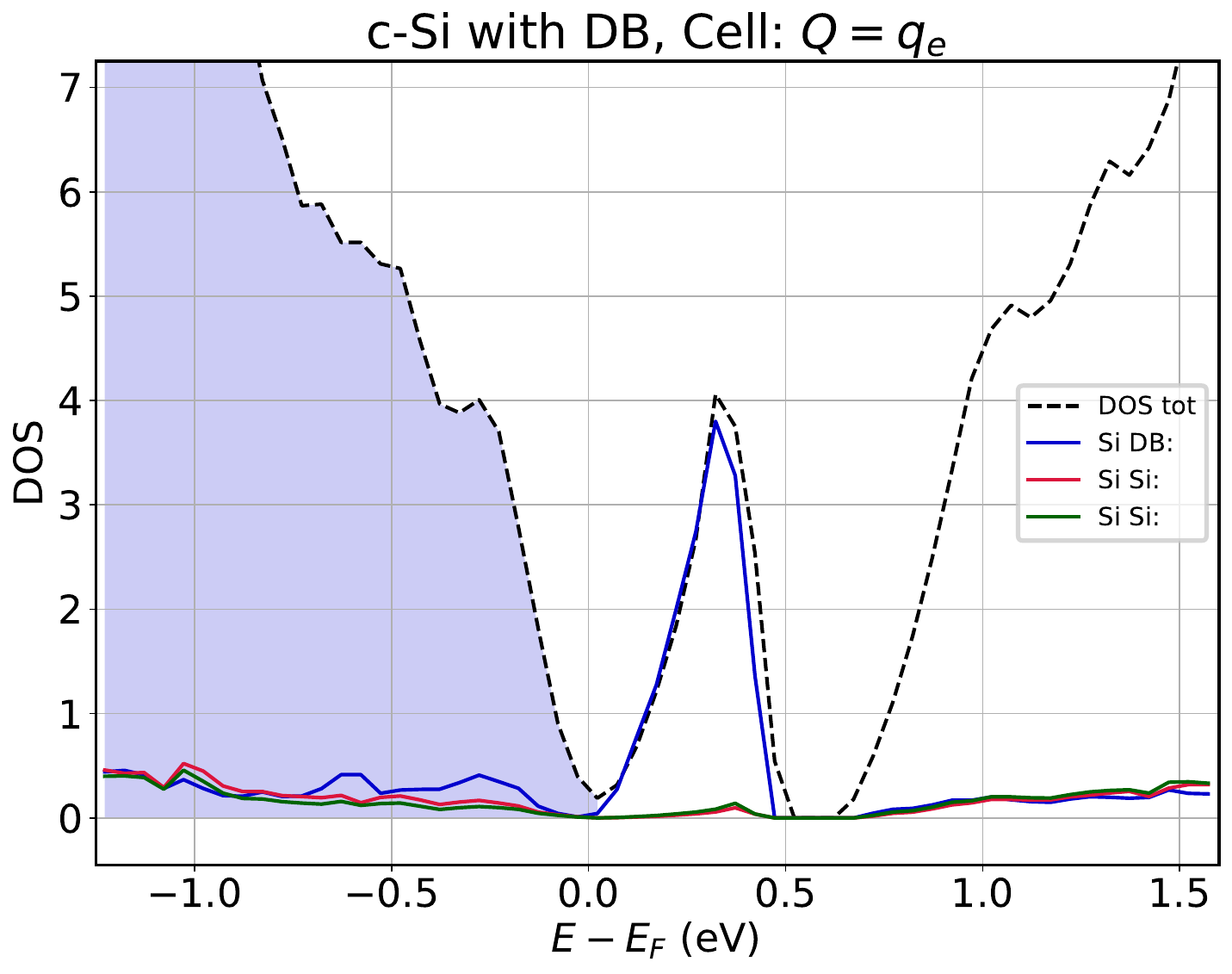}
    \label{fig:PQ_DB}
    \end{subfigure}
    \hfill
    \begin{subfigure}[t]{0.30\textwidth}
    \caption{}
    \vspace{0pt}
    \includegraphics[width =\textwidth]{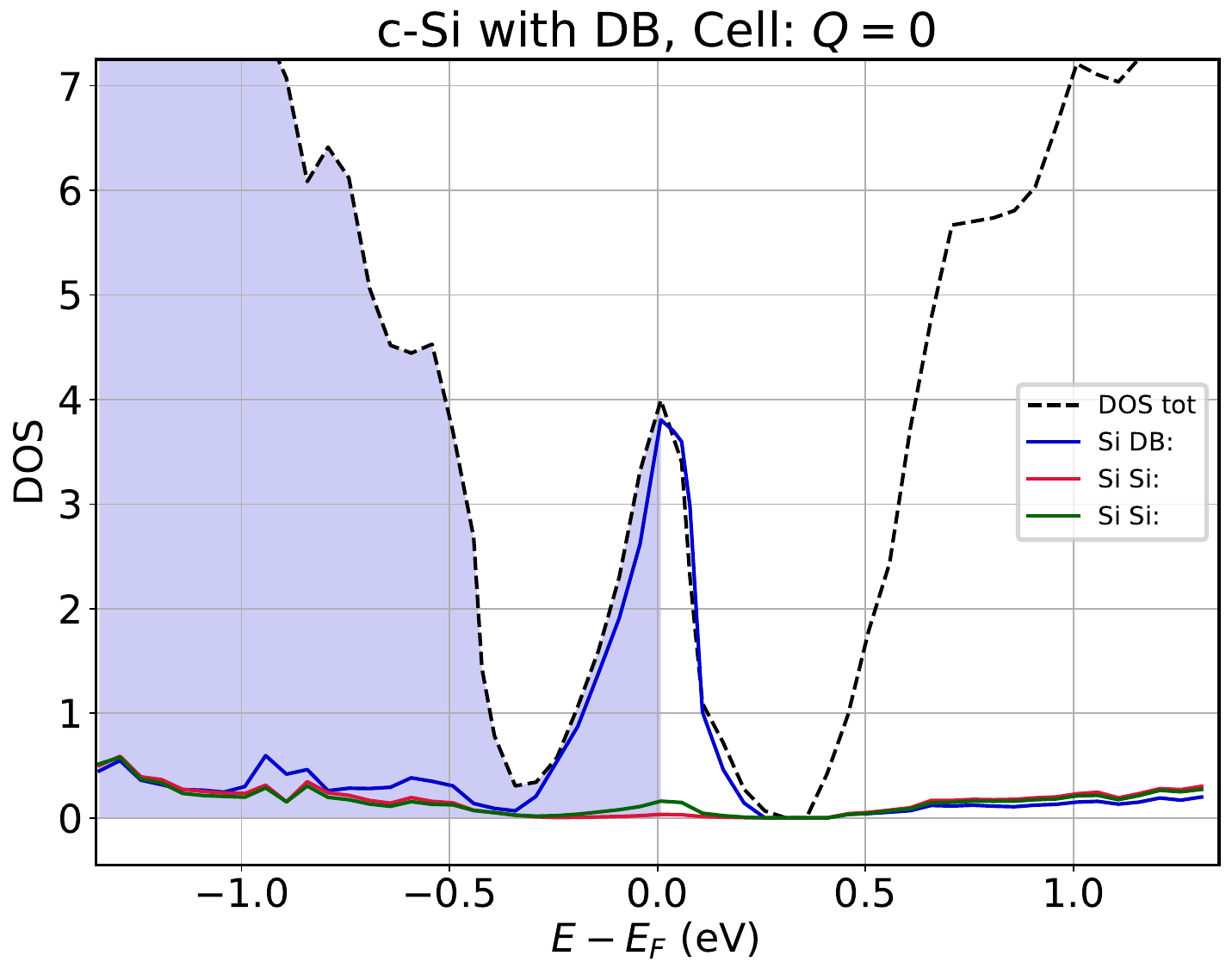}
    \label{fig:Q0_DB}
    \end{subfigure}
    \hfill
    \begin{subfigure}[t]{0.30\textwidth}
    \caption{}
    \vspace{0pt}
    \includegraphics[width =\textwidth]{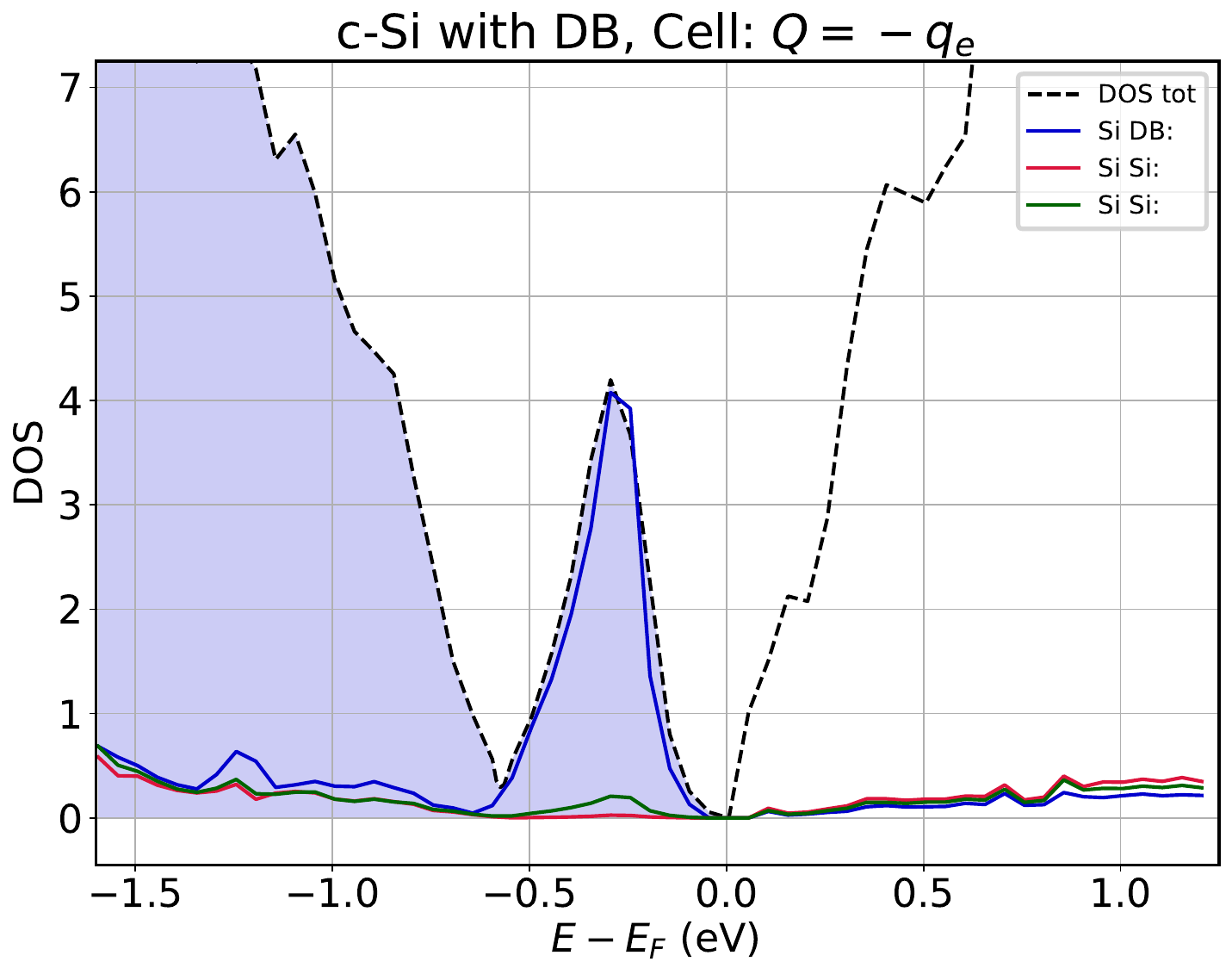}
    \label{fig:NQ_DB}
    \end{subfigure}
    \vspace{-10pt}
    \caption{The projected density of states for our three 63 Si + 3 H super cells with total cell charges  a) $Q=+q_e$, b) $Q=0$, and c) $Q=-q_e$. In the absence of a BCH complex the DB occupation is as expected.}
    \label{fig:DB_PDOS}
\end{figure}

We began our ``extra charge" analysis by exploring the occupation of the DB without a BCH complex present. To do this we started with the initial fully passivated cell, from which one H was removed, leaving behind a Si DB. We then relaxed three cells using a variable cell relaxation with the three cell charges, and computed the PDOS and the L{\"o}wdin population.  We found that without a BCH complex nearby, the occupations of the Si DB follow the expected Fermi-Dirac statistics, shown in Fig.~\ref{fig:DB_PDOS}. We also computed the effective dangling bond charge, $q_{DB}$, which is the difference between the L{\"o}wdin population of the Si DB and the average L{\"o}wdin population of the other Si in the super cell. We found that $q_{DB}$, reported in Table \ref{tab:qeff_DB}, tracked well with the occupations provided by the PDOS, granting us additional insight into the local electronic configurations.

\begin{table}[h]
    \centering
    \begin{tabular}{|c|c|}
    \hline
    Cell Charge & $q_{DB}$ \\ 
    \hline
         $+q_e$ & 0.149 \\
        \hline
         $0$ & -0.037 \\
        \hline
         $-q_e$ & -0.208 \\
        \hline    
    \end{tabular}
    \vspace{10pt}
    \caption{L{\"o}wdin population of Si DB for super cell charges $Q=q_e,0,-q_e$}
    \label{tab:qeff_DB}
\end{table}

After establishing the occupation of the DB without a BCH complex, we turned our attention to the electronic structure of a BCH complex without a Si DB. To do this we placed one H in otherwise perfect c-Si, then computed the PDOS and L{\"o}wdin population. We found that even in an otherwise perfect crystal the BCH complex creates a high energy state near the conduction band minimum. 

Here we point out that determining the precise energy of this state relative to the CBM in c-Si is problematic with the computational methods used in this work. A well known problem with semi-local DFT is the underestimation of the bandgap of many materials \cite{DFTdefects1}. For Si, DFT predicts the bandgap of $\sim 0.6$ eV compared to experimentally determined 1.13 eV. Thus, the placement of the  BCH state in the conduction band tail could be due to the under estimation of the bandgap. Additionally, experiments have shown the Si DB to have a $0/-$ charge transition of $\sim 0.2 $ eV, i.e. $U \sim 0.2$ eV. Our our study thus far did not include Hubbard corrections; therefore, the relative energy difference between the BCH state and the negative Si DB could be in error. To eliminate, to the best of our ability, any method based faulty results, we conducted a substantial additional computational effort including DFT/LSDA+U, DFT+U+V, and hybrid functional calculation with PBE0 \cite{lsda_U,Campo_2010dft_u_v,DFPT_U,PBE0}, details of which can be found in Supplementary Materials. These efforts were informative and supported our findings using standard DFT; however, the additional computational cost made continuing with these methods prohibitive. Therefore, we continued our study using the initial methods described earlier.

After validating our methods we were confident to conclude that: (1) a BCH complex creates a localized high energy state near the CBM, (2) this state is higher in energy than the energy states of the Si DB.

\begin{figure}[h]
     \centering
    \begin{subfigure}[t]{0.30\textwidth}
    \caption{}
    \vspace{0pt}
    \includegraphics[width = \textwidth]{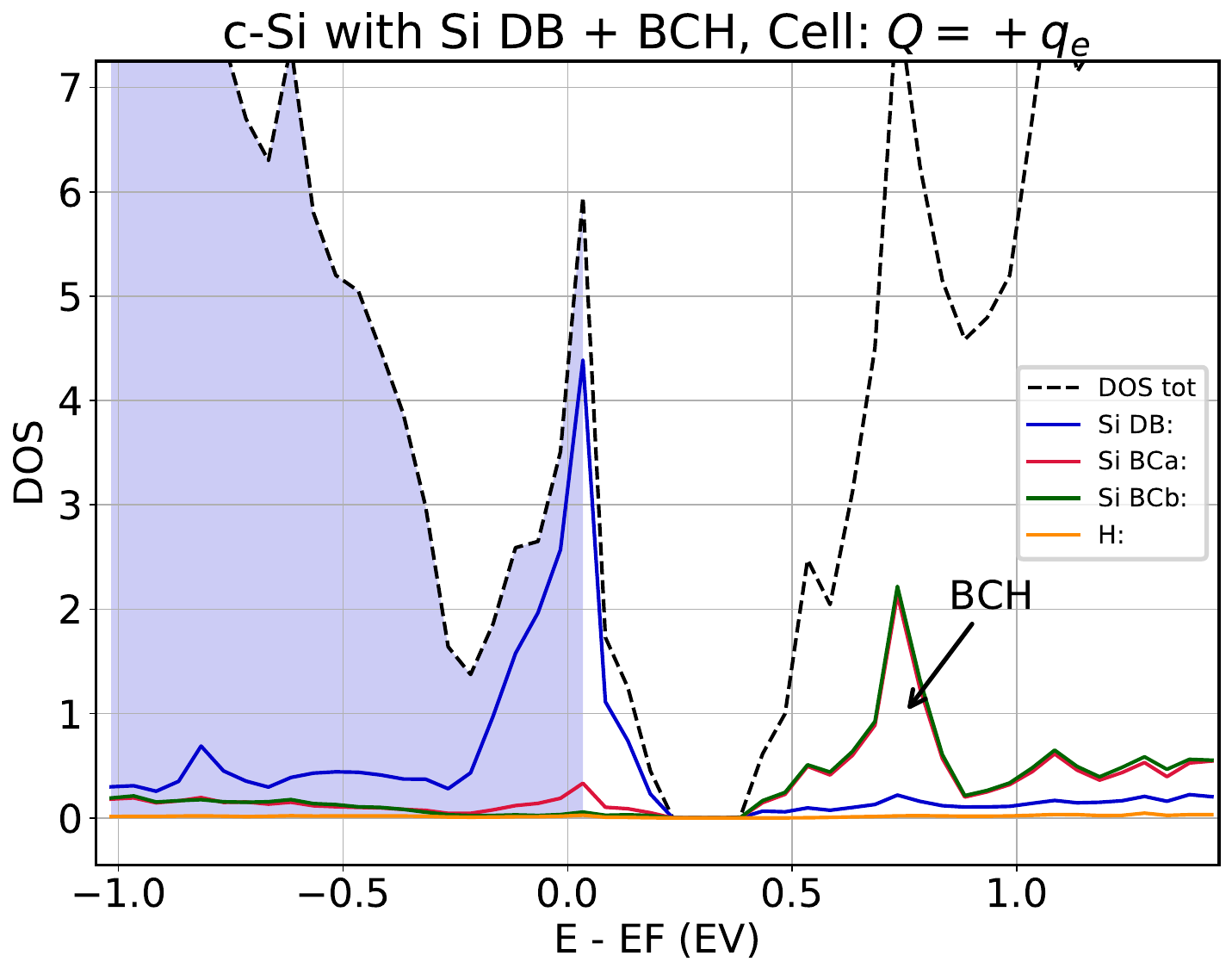}
    \label{fig:PQ_BCH}
    \end{subfigure}
    \hfill
    \begin{subfigure}[t]{0.30\textwidth}
    \caption{}
    \vspace{0pt}
    \includegraphics[width =\textwidth]{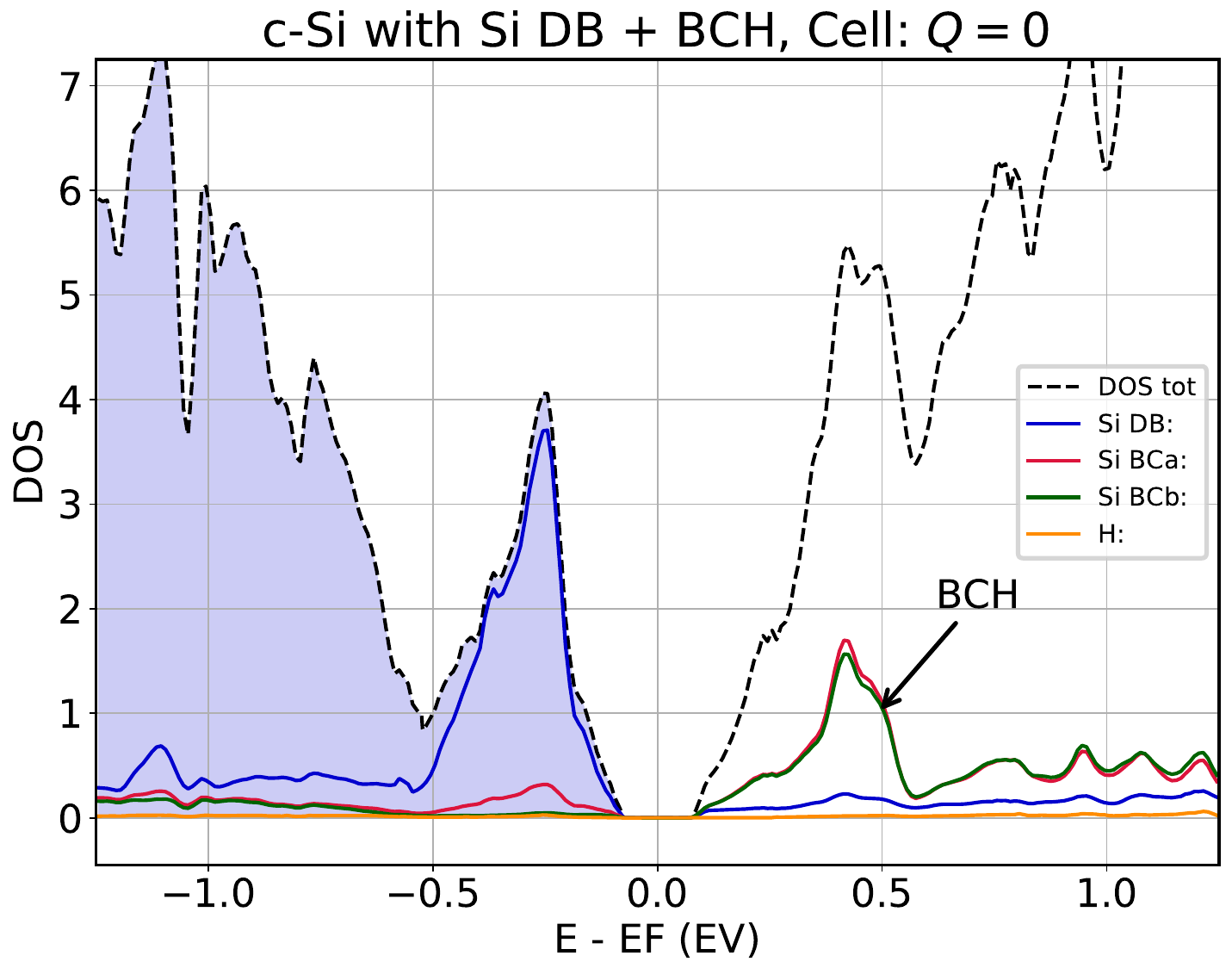}
    \label{fig:Q0_BCH}
    \end{subfigure}
    \hfill
    \begin{subfigure}[t]{0.30\textwidth}
    \caption{}
    \vspace{0pt}
    \includegraphics[width =\textwidth]{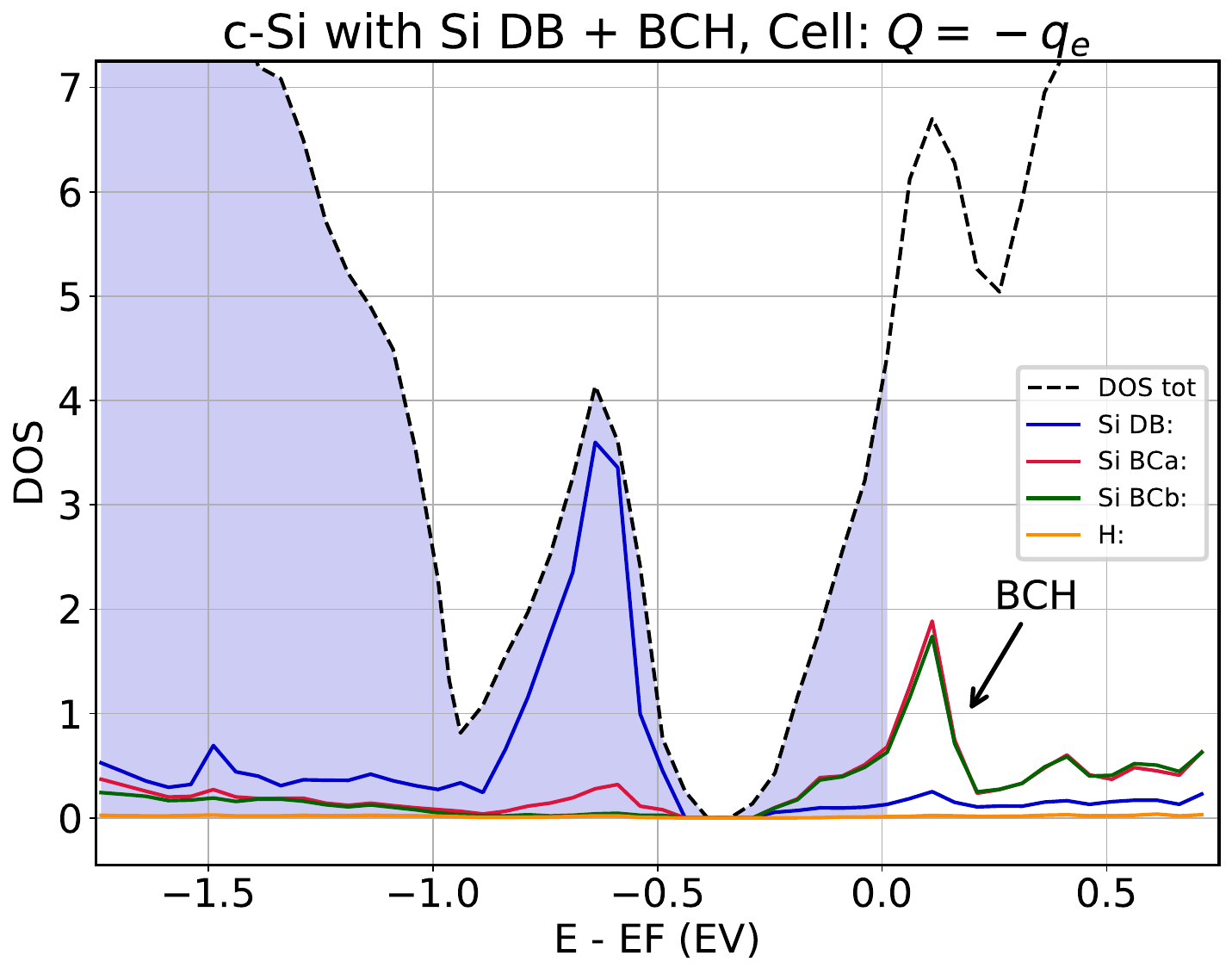}
    \label{fig:NQ_BCH}
    \end{subfigure}
    \vspace{-10pt}
    \caption{The projected density of states for c-Si with one Si DB and one BCH complex for three super cell charges: a) $Q=+q_e$, b) $Q=0$, and c) $Q=-q_e$. With an adjacent BCH complex, the occupation of the Si DB reflects that of the cell charge plus a donor.}
    \label{fig:BCH_PDOS}
\end{figure}

With our newfound insight we returned to our study of the three defect structures containing a Si DB and a BCH complex. For each of these cells, we computed the electronic properties using our same PDOS and the L{\"o}wdin population analysis. Figure \ref{fig:BCH_PDOS} shows the PDOS and Table \ref{tab:qeff_BCH} reports the $q_{eff}$ for these calculations, revealing our first key finding of this work.
We found that, in the vicinity of an available Si DB state, the BCH acts as a donor and an electron easily spills over into the lower energy state of the DB. Thus the occupation of the Si DB is now representative of the doping/cell charge, plus an additional donor. Further, the PDOS in Fig.~\ref{fig:BCH_PDOS} shows that the BCH state is primarily made up by the two BCH Si, $BC_a$ and $BC_b$, and the H barely contributes, if at all. Our calculated $q_{eff}$ corroborates this, which shows that in all thee cell charges the hydrogen atom in the BCH complex is essentially neutral, and the charge that has long been associated with an interstitial H atom is in fact that of the nearby Si, with a large portion localized on the two Si of the BCH complex. In the presence of an electric field a charged H would directly experience an additional force driving its motion. However, for the charged BCH complex the additional force will mostly be felt by the two BCH Si, which are not highly mobile. Our results showing a charged BCH complex are consistent with experimental results that found that changes in the mobility of H in an external electric field were not consistent with the picture of a charged hydrogen atom \cite{Santos-Johnson-Street-1992,Santos-Johnson-Street-1993}.   

\begin{table}[h]
    \centering
    \begin{tabular}{|c|c|c|c|c|c|}
    \hline
         Cell Charge, $Q$ & $q_{H}$ & $q_{db}$  & $q_{BCa}$ & $q_{BCb}$ & $q_{BCH}$  \\
        \hline
         $+q_e$ & -0.030 & -0.055 & 0.191 & 0.200 & 0.362 \\
         \hline
         0 & -0.039 & -0.177 &  0.184 & 0.192 & 0.338  \\
         \hline
         $-q_e$ & -0.034 & -0.186 &  0.146 & 0.151 & 0.263 \\
         \hline 
    \end{tabular}
    \vspace{10pt}
    \caption{$q_{eff}$ for the key atoms in the final defect state: the interstitial H, $q_H$, the Si DB, $q_{DB}$, the two Si in the BCH complex, $q_{BC_a}, q_{BC_b}$, and the BCH complex, $q_{BCH}$. Strikingly, $q_H$ remains almost constant and is essentially neutral for all cell charges. For cell charges $+q_e$ and 0,  $q_{DB}$ is occupied by the electron donated by the BCH complex. In all three cells $q_{BCH}$ carries a large positive charge; however, the change in $q_{BCH}$ between the intrinsic cell and the n-type cell is 0.075 $q_e$, which shows that a large amount of the additional $e$ is localized on the BCH, compared to the $\frac{1}{N} \sim 0.015$ for a delocalized electron.}
    \label{tab:qeff_BCH}
\end{table}

\section{Defect Generation Dynamics in Doped c-Si}

After determining the unique electronic structure resulting from  the interaction between a BCH complex and an adjacent Si DB, we returned to the initial focus for this work: the asymmetric enhancement in the mobility of hydrogen in doped a-Si:H and its connection to the observed degradation in silicon heterojunction stacks. The next step in our SolDeg method was to determine the energy barriers controlling defect generation. To do this, we again turned to the nudged elastic band method; specifically, we used c-NEB as implemented in QE.
For the initial image we used our fully passivated 63 Si + 4 H cells, and for the final image we used the final defect configuration cells containing a Si DB and a BCH. We used 7 total images, resulting in 5 intermediate images which created a path connecting the initial and final images. 

To find the MEP we took a three step approach. First, the 5 intermediate images were relaxed using BFGS optimization with no climbing images to a path tolerance of $0.1 ~\text{eV/\r{A}}$. If convergence was achieved, the resulting path was used to restart the NEB calculation, with the highest energy image set as a climbing image and now using a path tolerance of $0.09 ~\text{eV/\r{A}}$.\footnote{The small reduction in the path tolerance was needed to trigger the initial BFGS cycle.} Finally, if convergence was reached for the climbing image run, that path was used to restart the NEB calculation keeping the same climbing image but now using a path tolerance of $0.05 ~\text{eV/\r{A}}$. This full procedure was carried out individually for our p-type, intrinsic, and n-type cells. 

\begin{figure}[h]
     \centering
     \begin{subfigure}[t]{0.45\textwidth}
    \caption{}
    \vspace{20pt}
    \includegraphics[width = \textwidth]{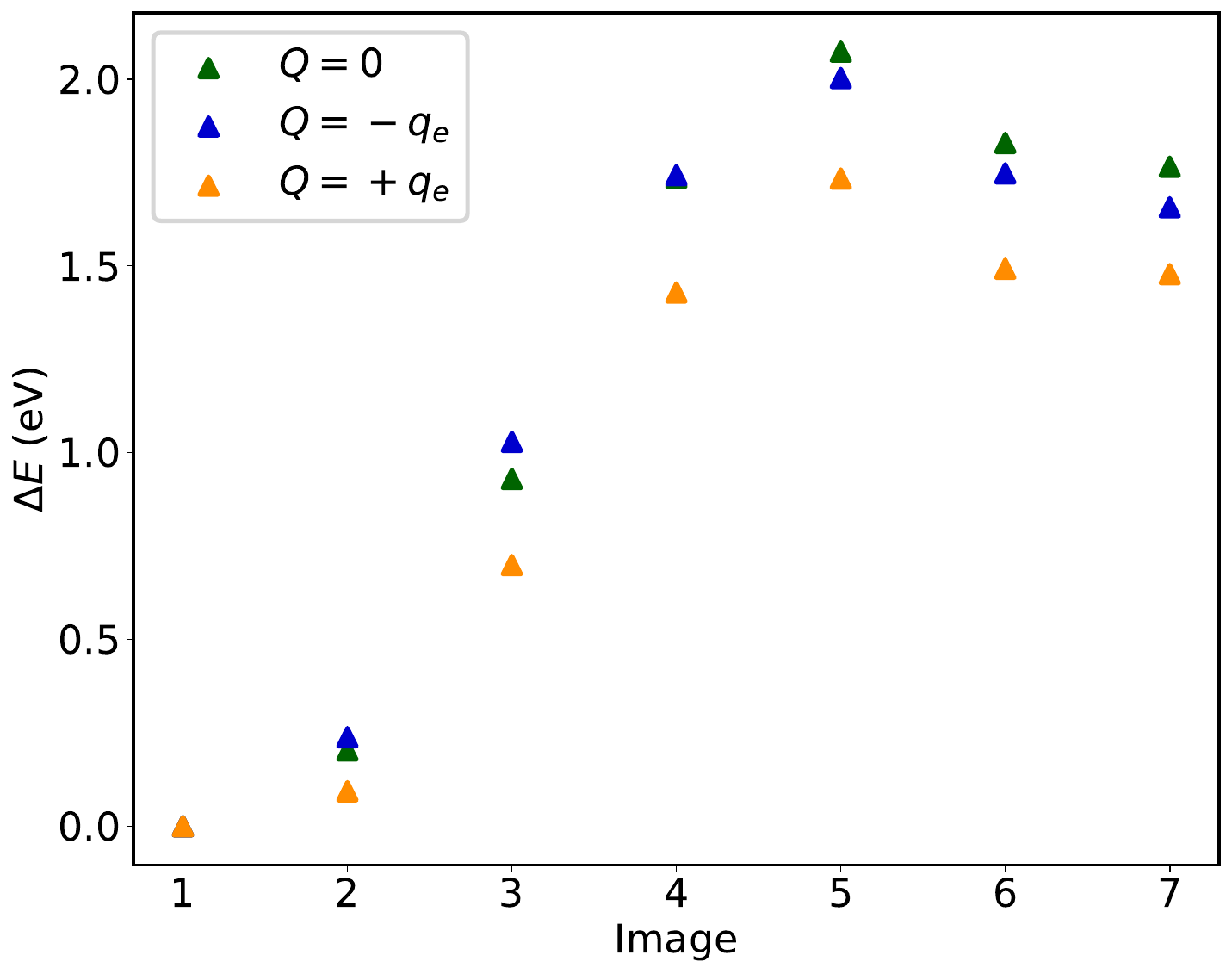}
    \label{fig:Fermi-MEP}
    \end{subfigure}
    \hfill
    \centering
     \begin{subfigure}[t]{0.45\textwidth}
    \caption{}
    \vspace{0pt}
    \includegraphics[width = \textwidth]{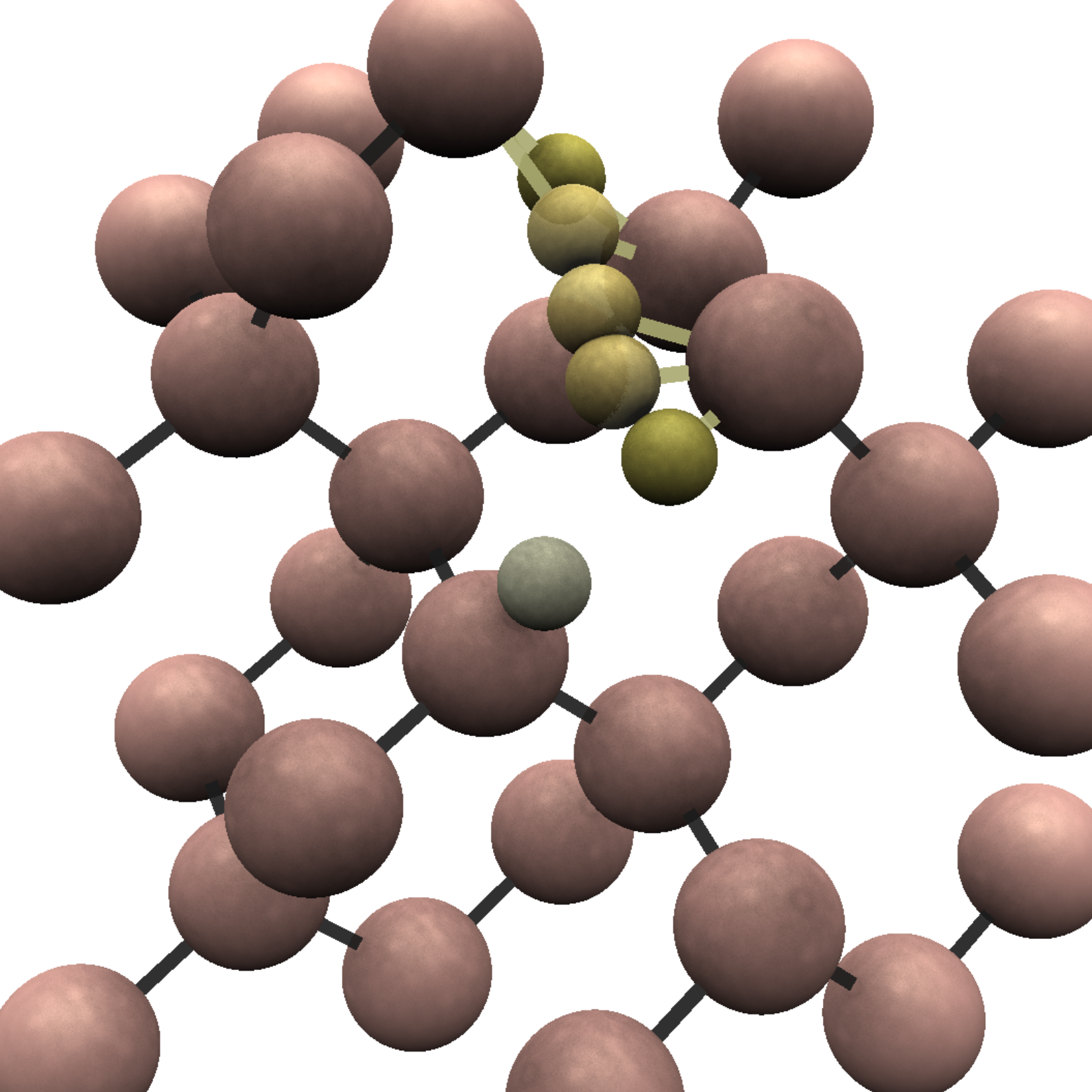}
    \label{fig:MEP-path}
    \end{subfigure}
    \begin{subfigure}[t]{0.5\textwidth}
    \caption{}
    \vspace{10pt}
    \hspace{10pt}
    \begin{tabular}{|c|c|c|}
    \hline
         Cell Charge & $E_{BB}$ (eV) & $E_{recap}$ (eV) \\
        \hline
        $+q_e$ & 1.73 & 0.26 \\ 
        \hline
        0 & 2.07 & 0.31 \\
        \hline
        $-q_e$ & 1.98 & 0.35 \\
        \hline
         
    \end{tabular}
    \label{tab:barriers}
    \end{subfigure}
    \vspace{5pt}
    \caption{\SubCap{a} Minimum energy paths energy barriers for Si-H bond breaking in our p-type, intrinsic, and n-type cells. \SubCap{b} Shows the corresponding atomic motion of the H along the MEP. \SubCap{c} Gives the corresponding values for the energy barriers. Remarkably we find the same asymmetric reductions for our doped cells as reported by experiment.}
    \label{fig:Fermi-Barriers}
\end{figure}

The computed barriers and the corresponding MEPs are reported in Table \ref{tab:barriers} and shown in Fig.~\ref{fig:Fermi-MEP} respectively. These findings represent the second key result of this work. We find that the computed energy barriers for our p-type, intrinsic, and n-type cells show the same asymmetric reduction as those predicted by experimental observations \cite{BEYER1991,Street-Kakalios-1987,Nickel-Beckers2002}.

\subsection{Zero Point Energy Correction}
According to TSS the rate of a reaction, $k^{\ddag}$, is given by the probability that a reactant $A$ reaches the TS multiplied by the flux through the TS
\begin{equation}
    k^{\ddag} =  \frac{\mathcal{Z}^{\ddag}}{\mathcal{Z}^{A}} v^{\ddag}
    \label{eqn:tst_1}
\end{equation}
\noindent
where $\mathcal{Z}^{\ddag}/\mathcal{Z}^{A}$ are the partition functions of the TSS/reactants, and $v^{\ddag}$ is the reaction coordinate velocity through the TS \cite{TST,Jech_Thesis,KOM}. 
For a quantum system near the ground state, an important correction to the $\mathcal{Z}^{\ddag}/\mathcal{Z}^{A}$ term in Eq.~\ref{eqn:tst_1} is the difference in the zero point energies of the thermal partition functions ($\Delta E_{ZP}$). $\Delta E_{ZP}$ is commonly computed using using the first order approximation
\begin{equation}
    \Delta E_{ZP} = \sum_n \frac{1}{2} \hbar \omega_n^{\ddag} - \sum_m \frac{1}{2} \hbar \omega_m^A
    \label{eqn:zpe}
\end{equation}
where $\omega^{\ddag}/\omega^A$ are the phonon frequencies at the transition state and initial state respectively. We computed the phonon spectrum for the initial, transition, and final state using \textit{phonon} package in QE, and then calculated Eq.~\ref{eqn:zpe} for both the forward and reverse barriers using all real frequencies. Due to the substantial computational expense, this was done for the intrinsic cell only; however, $\Delta E_{ZP}$ should be roughly similar in the doped cells. Table \ref{tab:ZPE} shows our computed $\Delta E_{ZP}$ and the corrected Si-H bond breaking barriers, which now align closely with previously reported values \cite{BEYER2003,Street-Kakalios-1987,SolDegH}.
\begin{table}[h]
    \centering
    \begin{tabular}{|c|c|c|}
    \hline
         Barrier & $\Delta E_{ZP}$ (eV) & Corrected Barrier (eV) \\
         \hline
         $E_{forward}$ & -0.632 & 1.44 \\
         \hline
         $E_{reverse}$ & 0.053 & 0.36 \\
        \hline
    \end{tabular}
\vspace{10pt}
    \caption{$\Delta E_{ZP}$ and the corrected energy barriers for Si-H bond breaking in intrinsic c-Si.}
    \label{tab:ZPE}
\end{table}

\subsection{The Origin of The Fermi Level Dependence of Si-H Bond Breaking}
After reproducing the experimentally observed asymmetric change in the Si-H bond breaking barrier with regard to the Fermi level, we proceeded to determine the underlying physics responsible for this phenomena. We recalled from our analysis of the BCH complex that the occupation of the Si DB showed a similar asymmetry, in that it was doubly occupied in both the intrinsic and n-type cells, while only singly occupied in the p-type cell. Motivated by these observations we proceeded to explore the occupations and localized charge density at the transition state. To do this, we computed the PDOS and $q_{eff}$ of the transition state structure, given by the saddle point configuration of the NEB path. 

\begin{table}[h]
    \centering
    \begin{tabular}{|c|c|c|c|c|c|c|}
    \hline
         Cell Charge & $q_{H}$ & $q_{DB}$  & $q_{BC_a}$ & $q_{BC_b}$ & $q_{BCH}$ & $\Delta E_C$ \\
        \hline
         $q_e$ & 0.085 & -0.029 & 0.109 & 0.048 & 0.231 & 0.0042\\
         \hline
         0 & 0.074 & -0.182 &  0.130 & 0.059 & 0.253 & 0.0068\\
         \hline
         $-q_e$ & 0.049 & -0.200 &  0.127 & 0.066 & 0.232 & 0.0051\\
         \hline

    \end{tabular}
    \vspace{10pt}
    \caption{$q_{eff}$ for the atoms locally involved in the Si-H bond breaking process: the H, $q_H$, the newly formed Si DB, $q_{DB}$, the two BCH Si , $q_{BCa}$ and $q_{BCb}$, and the whole BCH complex, $q_{BCH}$. $\Delta E_C$ is the local change in the Coulomb energy for these atoms at the transition state.}
    \label{tab:qeff_TS}
\end{table}

\begin{figure}
     \centering
    \begin{subfigure}[t]{0.30\textwidth}
    \caption{}
    \vspace{10pt}
    \includegraphics[width = \textwidth]{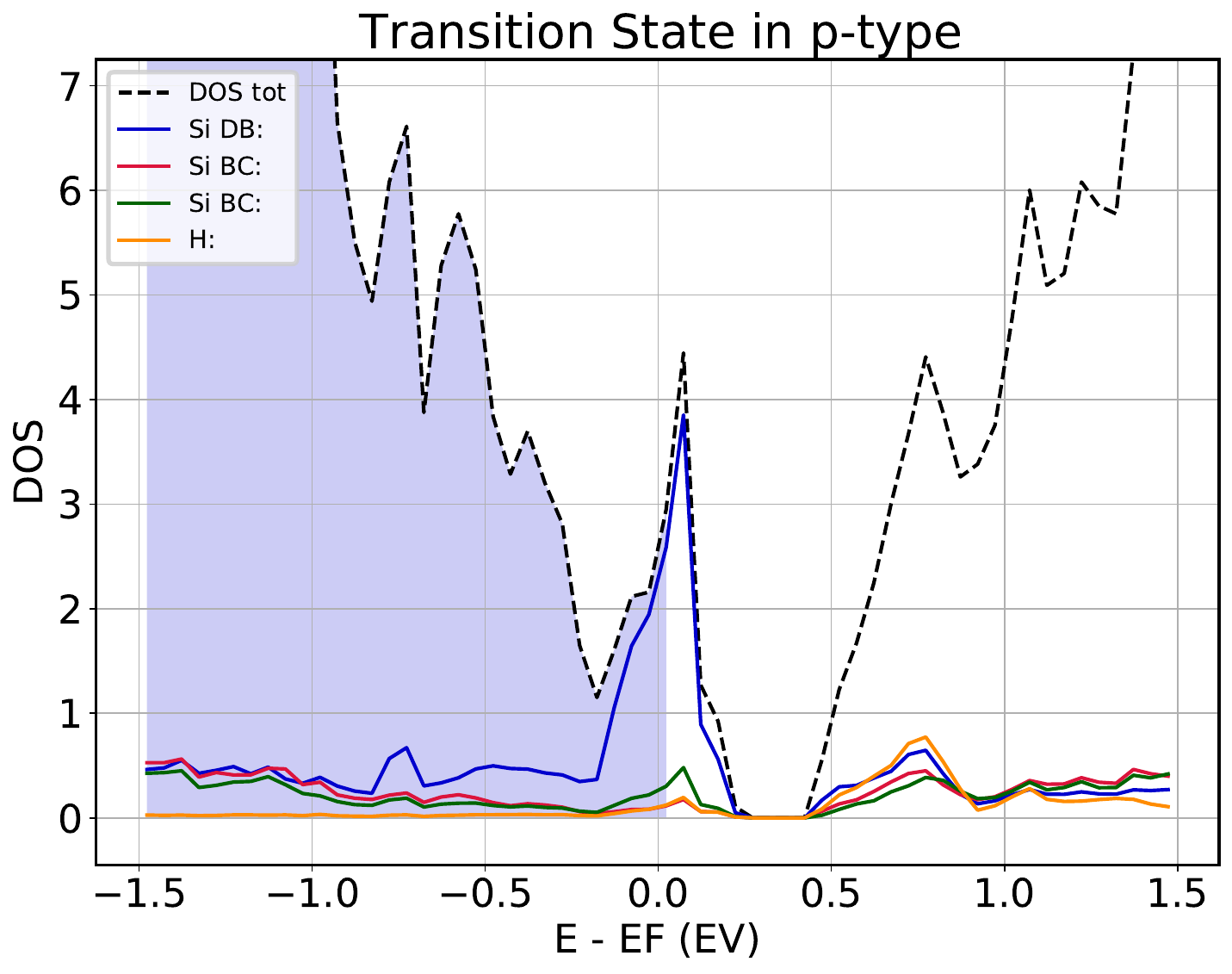}
    \label{fig:PQ_SP}
    \end{subfigure}
    ~
    \begin{subfigure}[t]{0.30\textwidth}
    \caption{}
    \vspace{10pt}
    \includegraphics[width =\textwidth]{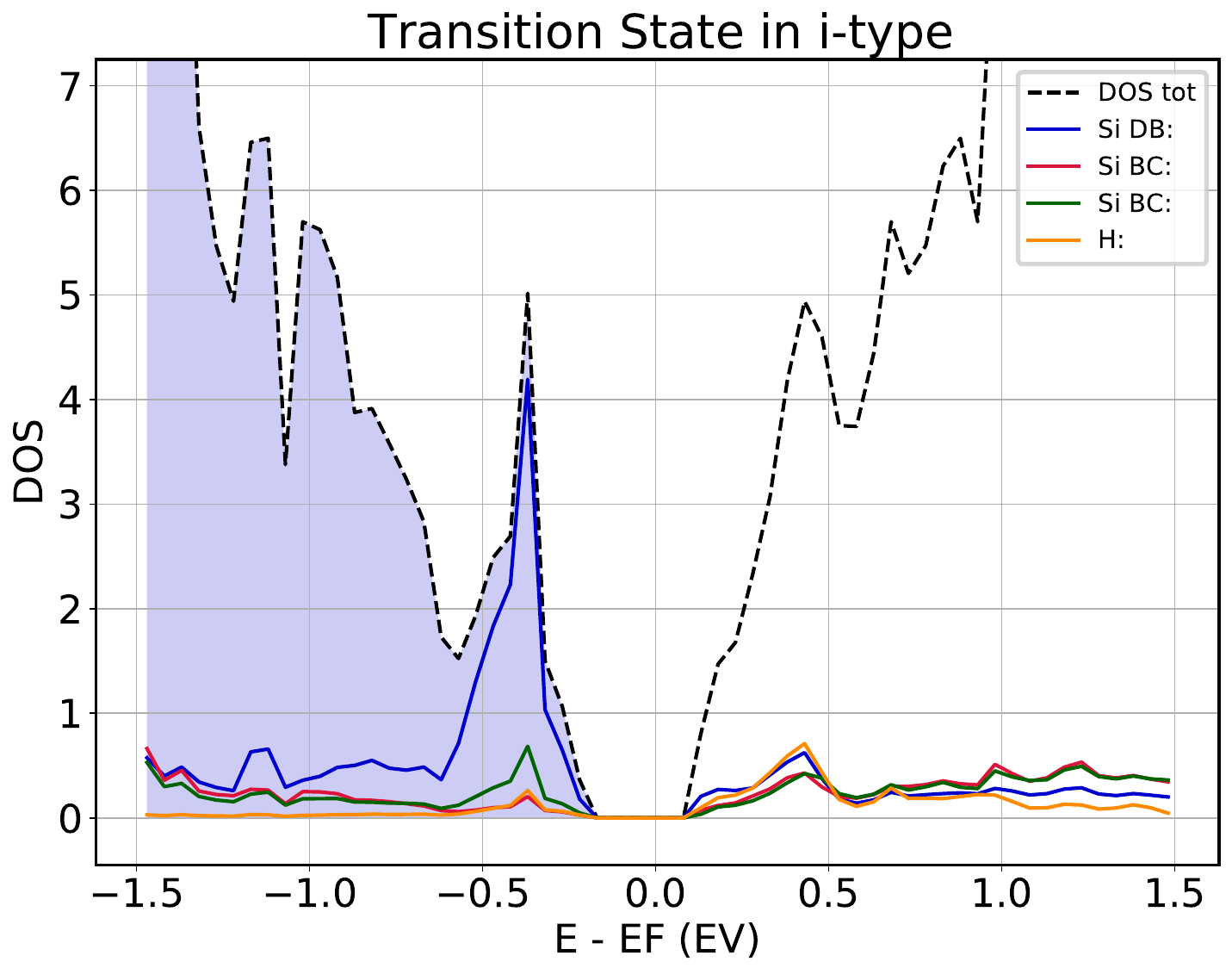}
    \label{fig:Q0_SP}
    \end{subfigure}
    ~
    \begin{subfigure}[t]{0.30\textwidth}
    \caption{}
    \vspace{10pt}
    \includegraphics[width =\textwidth]{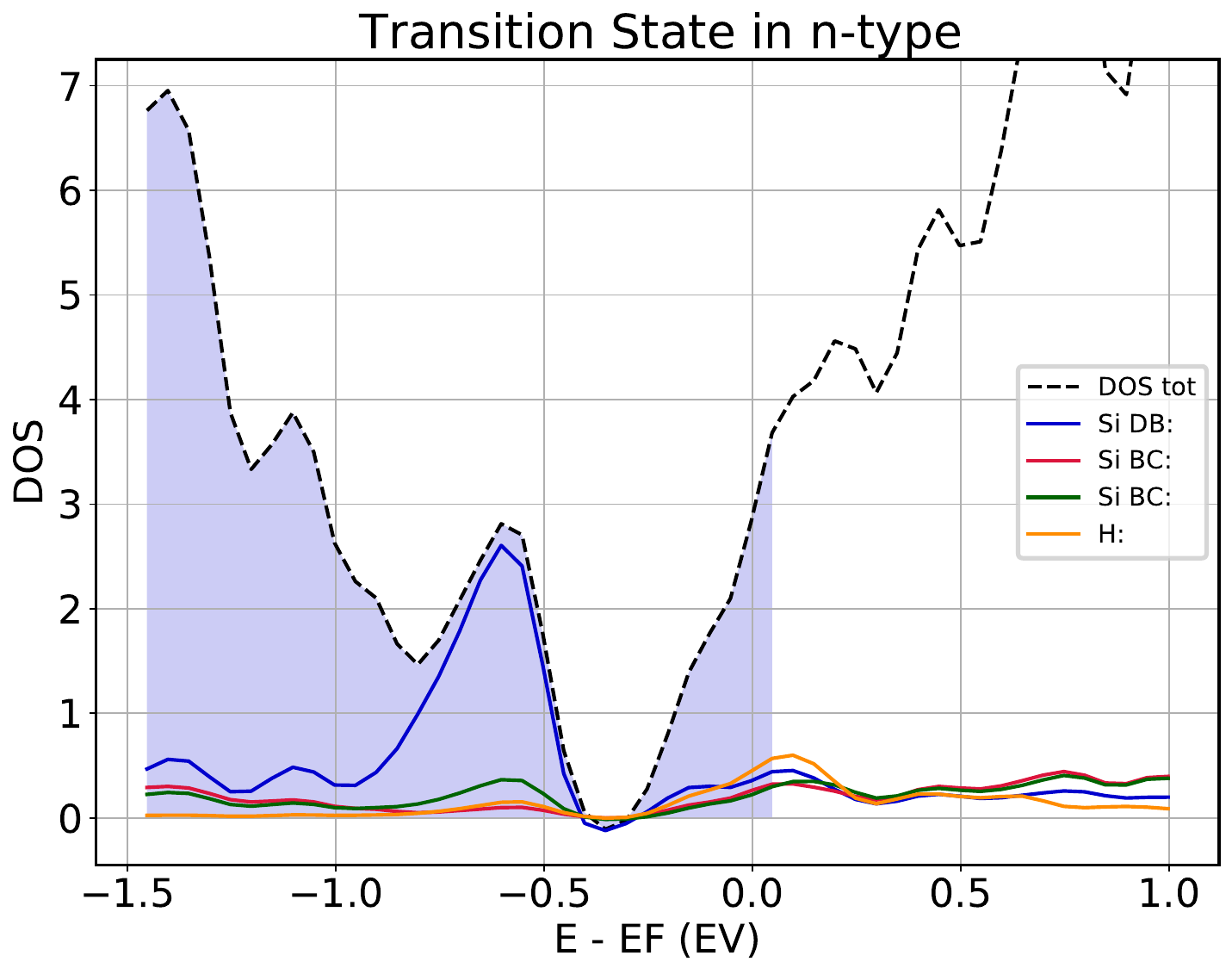}
    \label{fig:NQ_SP}
    \end{subfigure}
    \caption{The projected density of states at the transition state of the Si-H bond breaking process for a) p-type, b) intrinsic, and c) n-type super cells.}
    \label{fig:SP-pdos}
\end{figure}

The PDOS at the TS are shown in Fig.~\ref{fig:SP-pdos} and the corresponding $q_{eff}$ are reported in Table \ref{tab:qeff_TS}; together they represent our third key finding.
We found that at the TSS a precursor to the high energy state of the BCH complex had emerged. However, unlike the final BCH state, this state showed a substantial contribution from the H.
The emergence of this state created a high energy electron that spilled over into the lower energy state of Si DB, after which the occupation of the Si DB transitioned to extra donor filling observed in the final configurations. 
This analysis was further supported by our $q_{eff}$ calculations. Comparing the results presented in Table \ref{tab:qeff_TS} to those presented in Table \ref{tab:qeff_BCH}, we found that $q_{DB}$ showed very similar values at both the TSS and the final state. 
At the TSS, $q_{BC_a}$ is slightly lower than at the final state, and $q_{BC_b}$ is now close to neutral in contrast to its substantial positive charge shown at the final state. 
The most substantial difference seen was in $q_H$, which at the TSS is now positive and noticeably larger in the p-type and intrinsic cells than in the n-type cell. 
Combining these results we developed a hypothesis that the change in the Si-H bond breaking barrier was the result of changes in the local Coulomb interactions along the MEP. 
In the p-type cell the $q_H$ is largest but $q_{DB} \sim 0$, therefore the product $q_H *  q_{DB}$ is small. 
In the intrinsic cell $q_H$ is large and positive and $q_{DB}$ is large and negative, thus the product $q_H *  q_{DB}$ is large and negative, indicating a strong attraction. 
In the n-type cell $q_{DB}$ is large and negative but the additional electron reduces $q_H$, which reduces the product $q_H *  q_{DB}$, thus decreasing the attractive force. 
For a qualitative measure of this effect we determined the change in the Coulomb energy, $\Delta E_C$, between the initial image and the TSS image along the adiabat corresponding to $q_{eff}$ of the TSS. Specifically we computed
\begin{equation}
    \Delta E_C = q_H \bigg( \frac{q_{DB}}{\Delta r_{DB}} + \frac{q_{BC_a}}{\Delta r_{BC_a}} + \frac{q_{BC_b}}{\Delta r_{BC_b}} \bigg).
    \label{eqn:delC}
\end{equation}
$\Delta E_C$ for our three different Fermi levels is reported in Table \ref{tab:qeff_TS}, which shows a markedly similar asymmetry to the energy barriers. These findings support our model, where the Fermi level dependence of Si-H bond breaking is the result of changes in the local Coulombic interactions between the Si DB and the BCH complex.

\subsection{Interstitial Hydrogen Dynamics in Doped c-Si}

In our previous work we found that Si-H bond breaking was only one part of the total mechanism controlling hydrogen motion in Si \cite{SolDegH}. In that work we found that the ratios of the barriers controlling the recapture of a hydrogen by a Si DB, $E_{recap}$, the hopping out of the first Si BC into the next, $E_{drift}$, and the reverse of that process, $E_{return}$, had a substantial effect on the overall kinetics. To investigate similar effects we computed the energy barriers controlling the hopping of a H in a BCH to an adjacent Si-Si BC. This was done using the same NEB procedures outlined in the previous section. The computed ``next BC" hopping barriers, along with the Si-H bond breaking barriers, are reported in Table \ref{tab:full_barriers}. Using a set of coupled rate equations we computed an effective bond breaking barrier. The computed effective barriers showed only small variations from the Si-H bond breaking barriers. Thus, in absence of a density gradient the Si-H bond breaking barrier is a sufficient measure of the energy barrier controlling hydrogen kinetics, and thus defect generation, in crystalline and amorphous silicon. 


\begin{table}[h]
    \centering
   \begin{tabular}{|c|c|c|c|c|}
    \hline
         Super Cell Charge & $E_{BB}$ (eV) & $E_{recap}$ (eV) & $E_{drift}$ (eV) & $E_{return}$ (eV) \\
        \hline
        $+q_e$ & 1.73 & 0.26 & 0.26 & 0.36 \\ 
        \hline
        0 & 2.07 & 0.31 & 0.28 & 0.36 \\
        \hline
        $-q_e$ & 1.98 & 0.35 & 0.31 & 0.33 \\
        \hline
         
    \end{tabular}
    \vspace{10pt}
    \caption{Energy barriers determined by NEB for Si-H bond breaking and next BC hopping for our p-type, intrinsic, and n-type cells. There are slight differences in the next BC hopping barriers for the different cells; however, our computed effective barriers showed only slight variations from the Si-H bond breaking barrier. Therefore, the Si-H bond breaking barrier is a sufficient measure of the thermal barrier that controls H motion in amorphous and crystalline Si.}
    \label{tab:full_barriers}
\end{table}

\section{Defect Generation Dynamics in Doped a-Si:H}

After determining the mechanisms responsible for the changes in the Si-H bond breaking barrier, and therefore the barrier controlling hydrogen motion, in c-Si we proceeded to investigate how these findings might change due to the macroscopic differences, e.g., bandgap, and the microscopic differences, e.g., strained bonds, micro-voids, and other BCH, in a-Si:H. We began by creating a-Si:H super cells of roughly the same size to keep computational costs lower and to remain consistent with finite size effects such as localized charge-jellium interactions. 

 The a-Si:H structures were created with classical MD using the LAMMPS simulator \cite{LAMMPS} with our previously developed Si:H GAP interatomic potential \cite{SiHGAP}. We began by creating c-Si structures with randomly distributed vacancies and hydrogen atoms to achieve initial structures with densities of 2.28, 2.23, and 2.19 $\textrm{g}/\textrm{cm}^3$ with 15 at\% hydrogen. In accordance with our previous work, we used a validated melt-quench method, first heating the cell to 1800K to form liquid Si:H, followed by a quench to 1500K at a rate of $10^{13}\textrm{K}\textrm{s}^{-1}$. The cells were then allowed to equilibrate for 100 ps using canonical, NVT, sampling with a Nos{\'e}-Hoover thermostat. This was followed by a second quench to 500K at a rate of $10^{12}\textrm{K}\textrm{s}^{-1}$ using isothermal-isobaric, NPT, sampling with a Nos{\'e}-Hoover thermostat. Finally the cells were relaxed using a Hessian free truncated Newtonian optimization to a convergence tolerance of $10^{-3}$.
In some cells additional H were added to passivate Si DBs that interfered with with our desired calculations. Specifically, in some cells, two Si DBs that were spatially separate, i.e., not part of the same micro-void, were able to interact through the periodic boundaries and form a band, thus delocalizing the occupying electrons.
Prior to DFT calculations, all cells were relaxed with DFT using a variable cell relaxation with BFGS optimization to an energy tolerance of $10^{-4}$ Ry and a force tolerance of $10^{-3}$ Ry.

\begin{figure}
    \centering
    \includegraphics[width=0.45\linewidth]{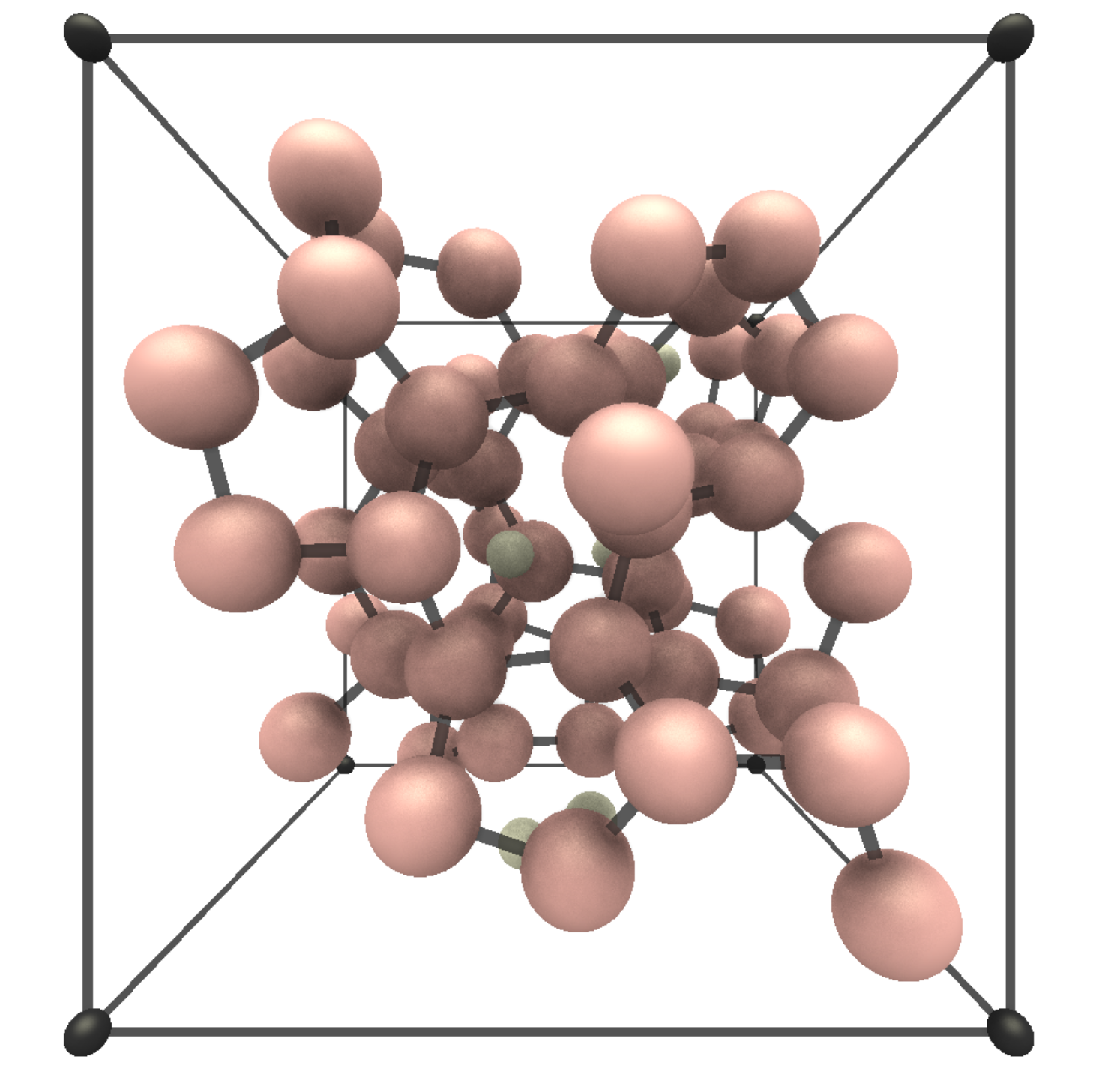}
    \caption{Example a-Si:H super cell containing 61 Si and 9 H, 15 at\% H.}
    \label{fig:asih_cell}
\end{figure}

\begin{figure}
     \centering
    \begin{subfigure}[t]{0.47\textwidth}
    \caption{}
    \vspace{0pt}
    \includegraphics[width = \textwidth]{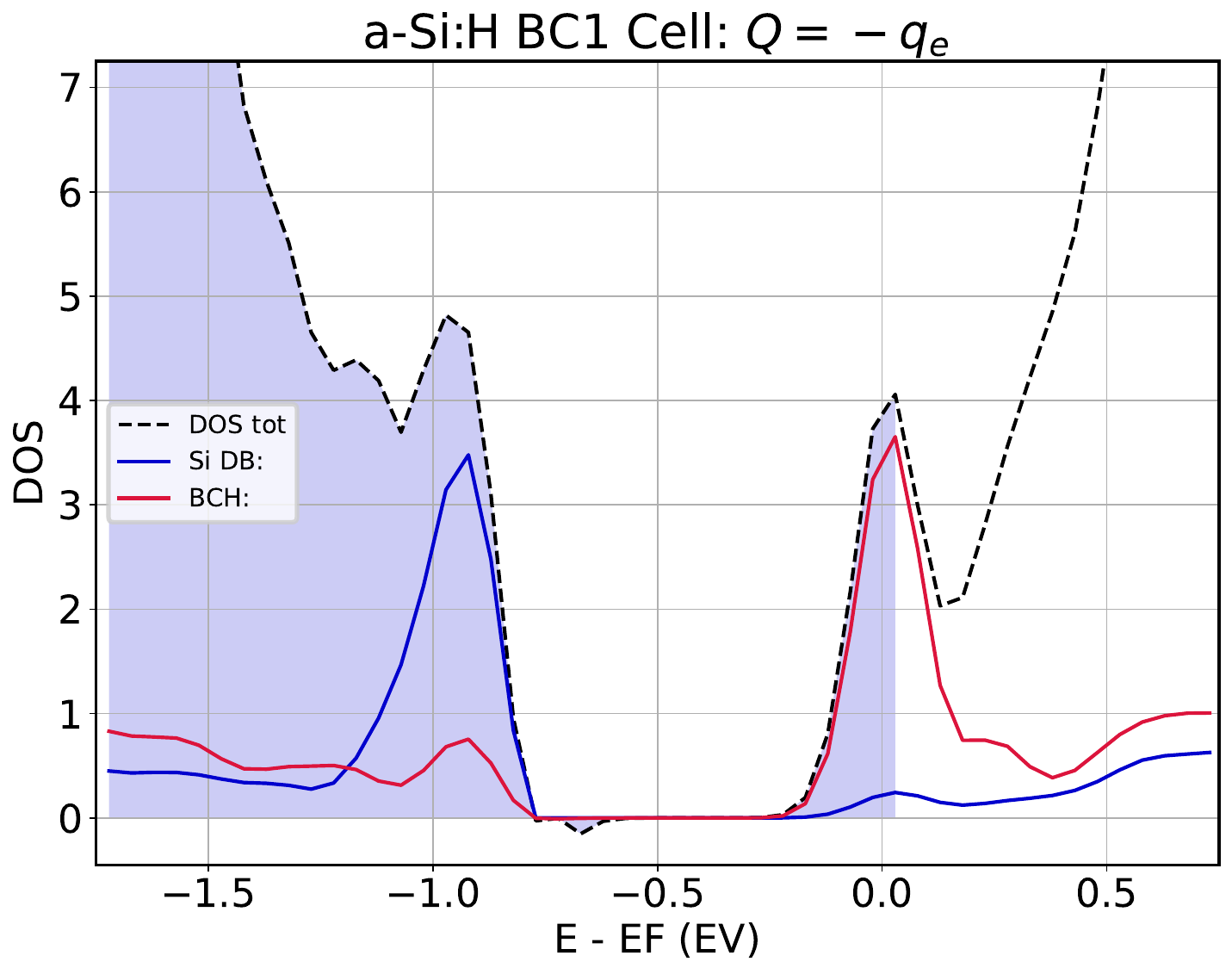}
    \label{fig:221_bc1}
    \end{subfigure}
    ~
    \begin{subfigure}[t]{0.47\textwidth}
    \caption{}
    \vspace{0pt}
    \includegraphics[width =\textwidth]{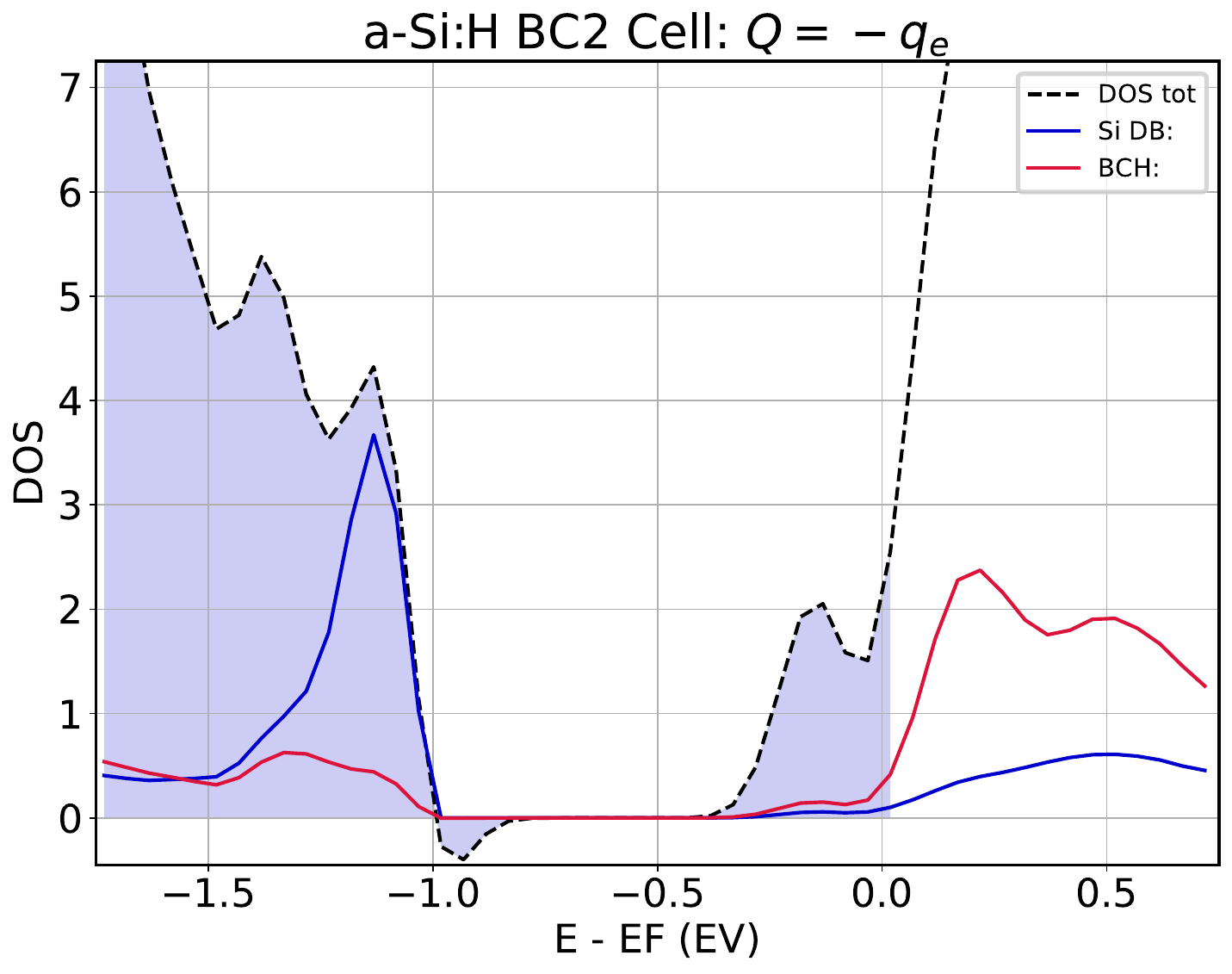}
    \label{fig:221_bc2_pdos}
    \end{subfigure}
    \vspace{-10pt}
    \caption{The projected density of states for the Si DB and the combined BCH complex in n-type a-Si:H with two different final defect structures. \SubCap{a} The BCH complex state falls below the conduction band. This allows the extra charge to better stabilize the charged BCH complex at the TSS thus further reducing the Si-H bond breaking barrier. \SubCap{b} The BCH complex state is completely within the conduction band and the barrier reduction is lost all together.}
    \label{fig:221_pdos}
\end{figure}

We carried out the same SolDeg procedure as was used for the c-Si: We created defects by displacing a H from an Si-H bond, computed the barriers controlling defect creation, then analyzed the electronic properties of the initial, transition state, and final structures. 
This was also done for doped, compensated, and charged cells. 
The results of our a-Si:H study were for the most part consistent with our results in c-Si. There were, however, some unique results that needed to be considered to construct a reliable model for Si-H bond breaking in a-Si:H. 
We found that in all cases where a hole was present at an energy lower than the Si DB, a similar $\sim 0.2-0.3$ eV reduction as was found in our p-type c-Si cell occurred in the Si-H bond breaking barrier. 
However, due to the variability of bonding configurations in the a-Si:H, the energy of the Si DB sometimes fell into the valence band tail and the barrier reduction was diminished or lost altogether. 
In the n-type a-Si:H cells a similar effect was seen, regarding the energy state of the BCH complex and the conduction band. 
Figure \ref{fig:221_pdos} shows the PDOS for two different final BCH configurations in our n-type a-Si:H cells. In Fig \ref{fig:221_bc1} the BCH complex state lies in the bandgap, and thus is highly localized. With this configuration, at the TSS the extra electron is much more localized on the BCH precursor state, which further reduces the barrier. 
In some case we found the Si-H bond breaking barrier was lowered by as much as 0.15 eV compared to 0.8 eV observed in the c-Si. On the other hand, Fig.~\ref{fig:221_bc2_pdos} shows that sometimes the BCH complex state was located completely within the conduction band and the barrier reduction was lost all together.

\begin{figure}
     \centering
    \begin{subfigure}[t]{0.40\textwidth}
    \caption{}
    \vspace{20pt}
    \includegraphics[width = \textwidth]{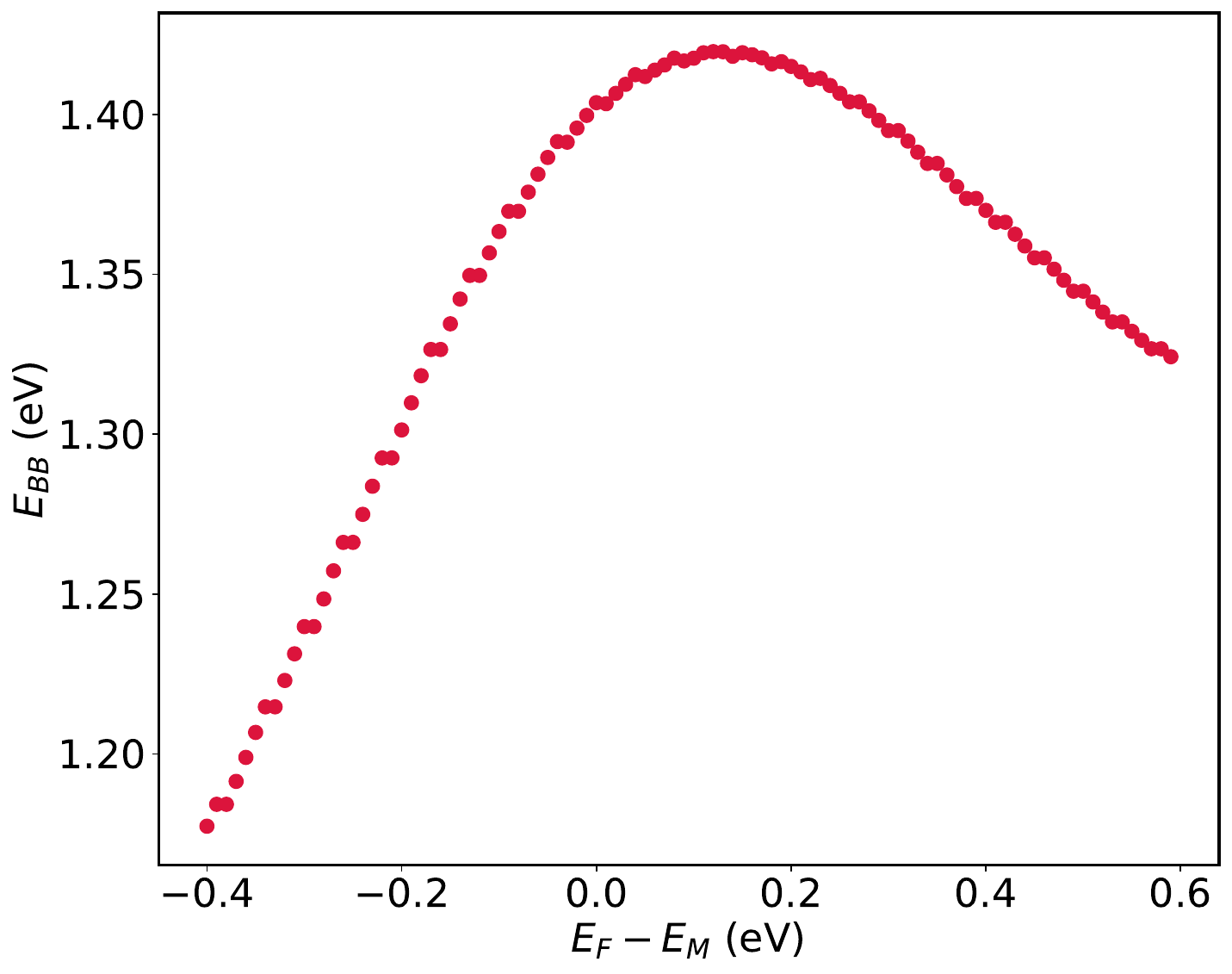}
    \label{fig:Soldeg-EBB}
    \end{subfigure}
    ~
    \begin{subfigure}[t]{0.47\textwidth}
    \caption{}
    \vspace{0pt}
    \includegraphics[width =\textwidth]{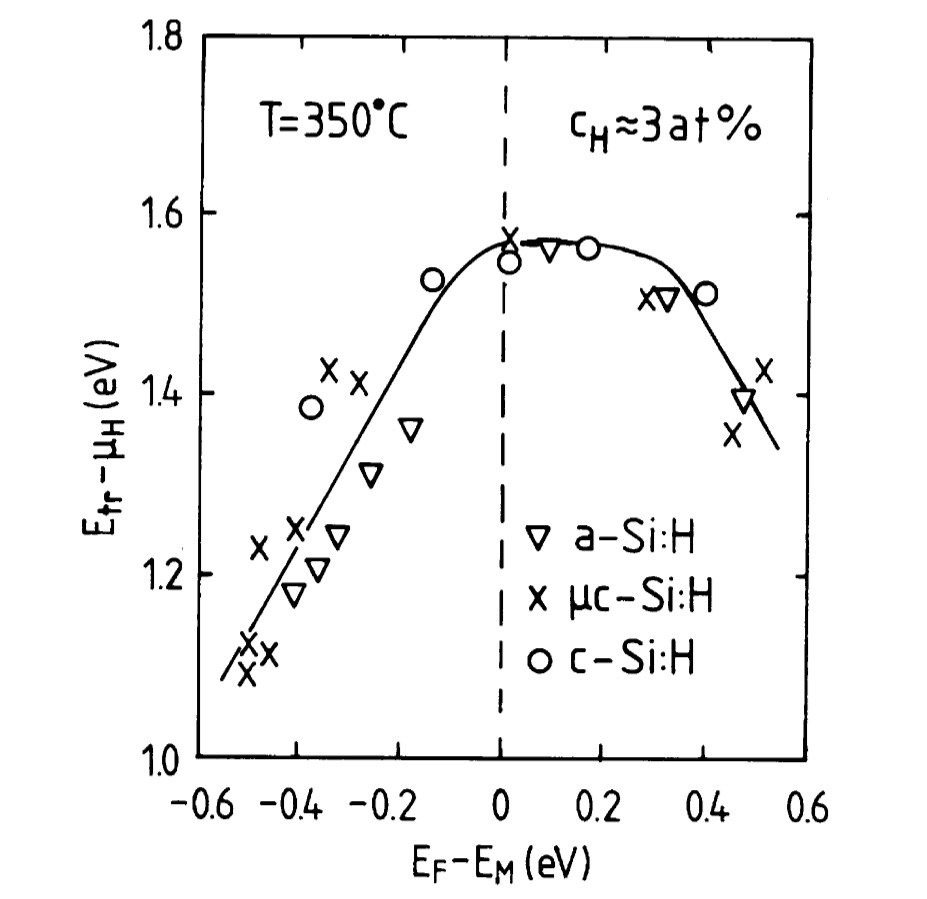}
    \label{fig:Beyer-EBB}
    \end{subfigure}
    \vspace{-10pt}
    \caption{\SubCap{a} The average Si-H bond breaking barrier as a function of the Fermi level relative to midgap determined by this work, side by side with \SubCap{b} the transport energy barrier, $E_{tr}$, for H diffusion in doped a-Si:H ($\triangle$), c-Si ($\circ$), and $\mu c$-Si ($X$). Reproduced from the legendary W. Beyer \textit{Solar Energy Materials \& Solar Cells}, \textbf{78} (2003).}
    \label{fig:EBB}
\end{figure}

These effects are caused by the distribution of bonding configurations in the a-Si:H, and thus the energy states of the Si DB and the BCH complex will be similarly distributed. 
Thus, at a given Fermi level the distributions of the energy states will have different occupations resulting in different bond breaking barriers. 
To model these effects requires distributions of energy levels for both Si DB and BCH in a-Si:H. 
Collecting large enough data sets to build up distributions was cost- and time-prohibitive; however, in our previous work we showed that the Si-Si and Si-H bond lengths \cite{soldeg}, as well as the Si-H bond breaking barriers \cite{SolDegH} were both Gaussian distributed. 
It is therefore reasonable to expect that the energy positions of the DB and BCH are also Gaussian distributed.
Using the mean values of the energy levels relative to the valence band maximum for the Si DB, $\bar{E}_{DB} = 0.4$ eV, and the BCH complex, $\bar{E}_{BCH} = 0.85$ eV, obtained in this work,  we constructed Gaussian distributed $E_{DB}$ and $E_{BCH}$ that were used to compute the average Si-H bond breaking barrier as a function of the Fermi level, using our NEB computed barriers with the ZPE correction. 
Figure \ref{fig:EBB} shows the results of our SolDeg-modeled effective bond breaking barrier next to the experimentally determined barriers controlling hydrogen diffusion in c-Si, $\mu c$-Si, and a-Si:H \cite{BEYER2003} as a function of the Fermi level. 
The compelling match between our model and experiment represents the fourth key finding of this work. The results of this work provide a consistent theory for the observed Fermi level dependence of hydrogen mobility and defect generation in a-Si:H. 
 
\section{Light-Induced Defect Generation in c-Si}

After determining the mechanism responsible for the Fermi level dependence of Si-H bond breaking in silicon and developing a model for $E_{BB}(E_F)$ in a-Si:H that closely matched with experimental results, we turned our focus to the effects of photo-excited carriers on hydrogen mobility and defect generation in c-Si and a-Si:H. 

\begin{figure}
     \centering
    \begin{subfigure}[t]{0.30\textwidth}
    \caption{}
    \vspace{10pt}
    \includegraphics[width = \textwidth]{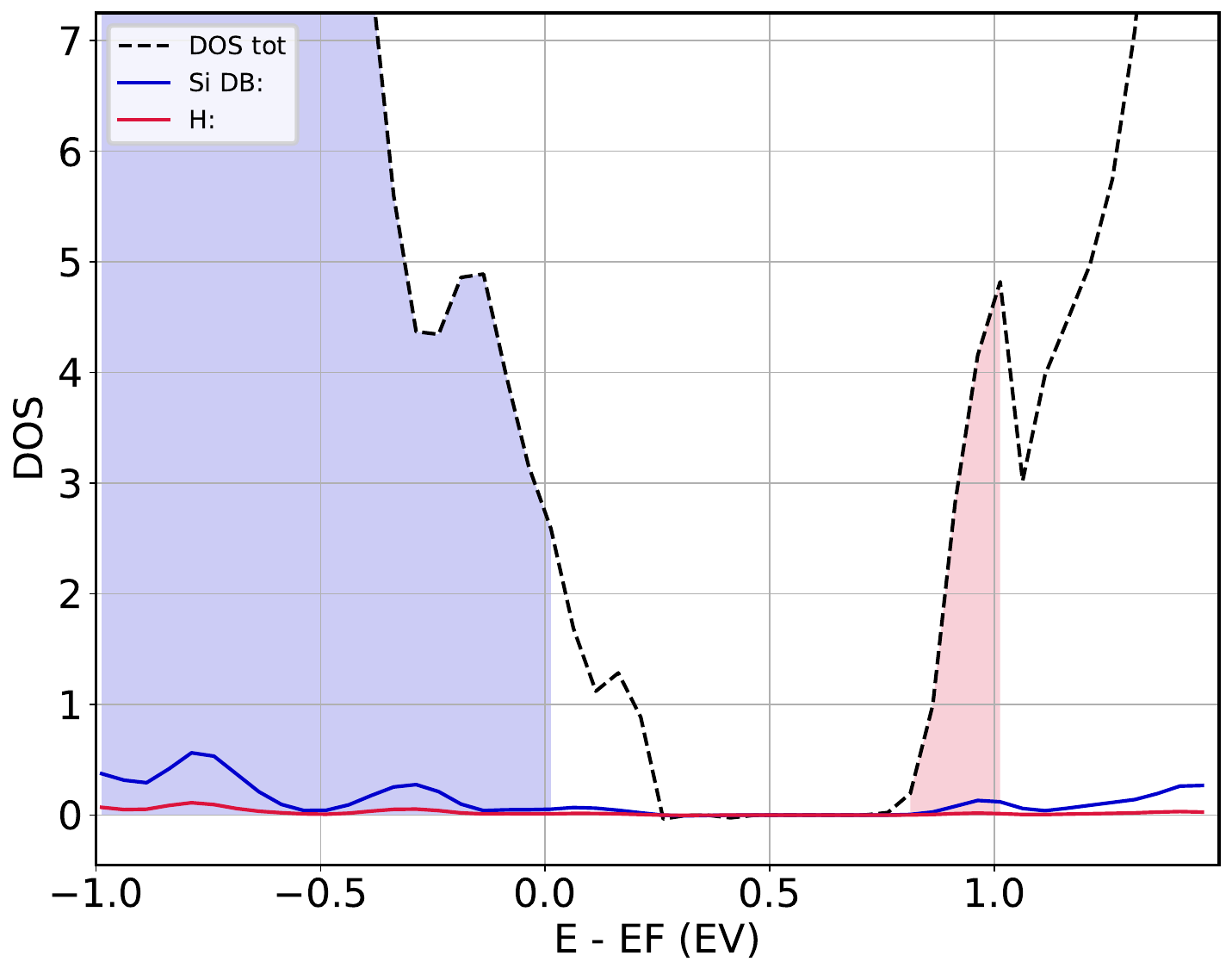}
    \label{fig:TC_IM1}
    \end{subfigure}
    ~
    \begin{subfigure}[t]{0.30\textwidth}
    \caption{}
    \vspace{10pt}
    \includegraphics[width =\textwidth]{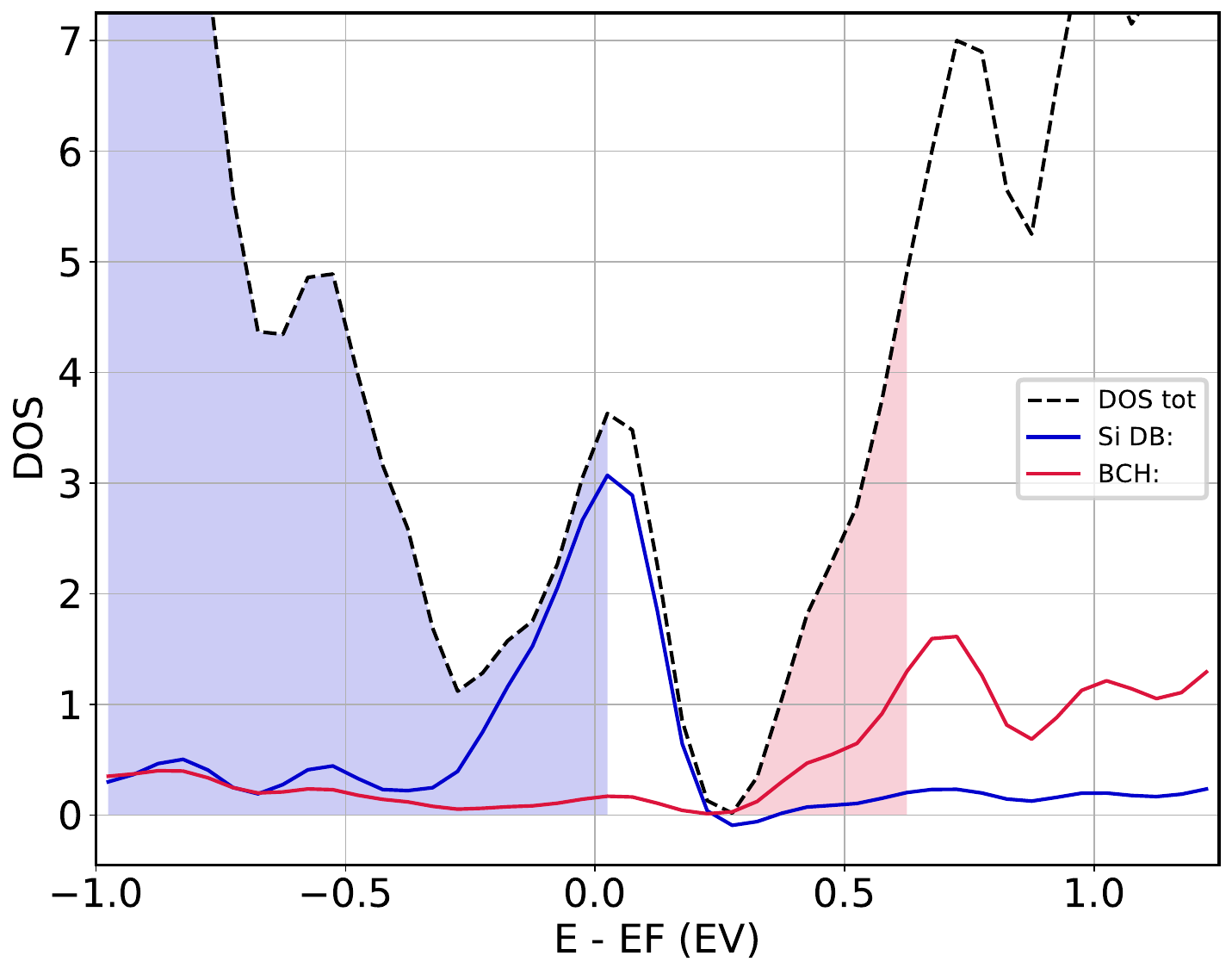}
    \label{fig:TC_SP}
    \end{subfigure}
    ~
    \begin{subfigure}[t]{0.30\textwidth}
    \caption{}
    \vspace{10pt}
    \includegraphics[width =\textwidth]{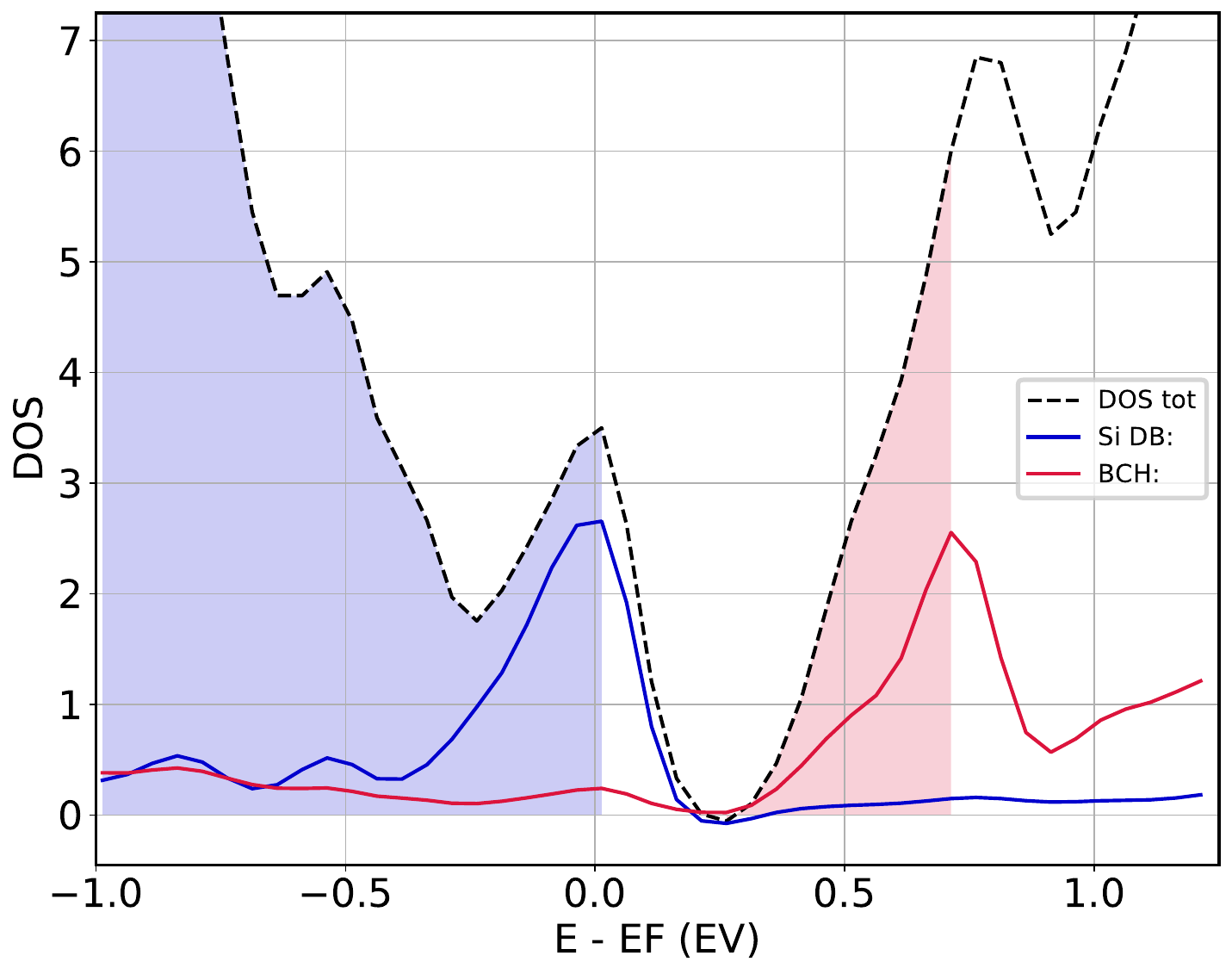}
    \label{fig:TC_BCH}
    \end{subfigure}
    \vspace{-10pt}
    \caption{ The projected density of states for 63 Si + 4 H super cell for \SubCap{a} the fully passivated initial image, \SubCap{b} the transition state image, and \SubCap{c} the final defect structure with a Si DB and a BCH complex for our illuminated Si-H bond breaking simulation. To simulate the effects of light we used the two chemical potential method to constrain one electron from the valence band to the conduction band. The \textcolor{blue}{blue fill} shows the occupied valence states, and the \textcolor{red}{red fill} shows the occupied conduction states.}
    \label{fig:TC-PDOS}
\end{figure}

To do this we used constrained density functional perturbation theory, (c-DFPT), to perform two chemical potential calculations that simulate photoexcited carriers by removing n electrons from the valence manifold and constraining them to occupy the conduction manifold, following the work of Marini and Calandra \cite{twoChem}. 
Figure \ref{fig:TC-PDOS} shows the PDOS for our intrinsic cell in the final Si DB + BCH defect state using the two chemical potential method. 
As shown by the blue fill, there is a hole in the valence states, accompanied by an electron, red fill, in the conduction states. 
After developing protocols for the two chemical potential calculations that reliably provided our desired electronic occupations, we proceeded to compute the Si-H bond breaking barriers for n-type, intrinsic, and p-type cells under illumination. 
For these calculations the cell charge method did not work because the valence manifold was not computed correctly. 
To carry out calculations of doped c-Si under illumination, we replaced one of the Si atoms in our fully passivated 63 Si + 4 H super cell with either a B or a P, thereby giving us p-type, intrinsic, and n-type cells. 
Each of these cells was relaxed using a variable cell relaxation with two chemical potentials, constraining one electron in the p-type and intrinsic cells, and two electrons in the n-type cell, into the conduction manifold\footnote{The n-type cell needed two electrons constrained to the conduction manifold to properly simulate illumination. Constraining only one electron to the conduction manifold in the n-type cell simulates the ionized P atom, which is not representative of illumination}. 
Using the relaxed cells we displaced one H to a next nearest neighbor Si-Si BC, followed by a fixed cell relaxation.
We then proceeded to compute the Si-H bond breaking barrier for these cells, using NEB combined with simulated illumination.
We again used a chain of 7 images with the fully passivated cell as the initial image and the Si DB + BCH complex as the final image. The intermediate images were relaxed to a path tolerance of 0.1 $\text{eV/ \r{A}}$. Table \ref{tab:dark-light} provides the results of these calculations along with our earlier ``dark" results for comparison. 
We find a substantial drop between 0.3-0.4 eV for all the Si-H bond breaking barriers, compared to their dark values. The reduction of 0.47 eV for the intrinsic barrier is in strong agreement with previously reported values \cite{Santos-Johnson-Street-1991}.

\begin{table}[h]
    \centering
    \begin{tabular}{|c|c|c|c|c|}
   \multicolumn{1}{c}{} &\multicolumn{2}{c}{Dark} & \multicolumn{2}{c}{Light} \\ \hline
         Type & $E_{BB}$ (eV) & $E_{recap}$ (eV) & $E_{BB}$ (eV) & $E_{recap}$ (eV) \\
        \hline
        p-type & 1.73 & 0.26 &  1.41 & 0.23 \\ 
        \hline
        intrinsic & 2.02 & 0.31 & 1.60 & 0.23 \\
        \hline
        n-type & 1.93 & 0.35  & 1.58 & 0.25 \\
        \hline
         
    \end{tabular}
    \vspace{10pt}
    \caption{Energy barriers for Si-H bond breaking in our p-type, intrinsic, and n-type cells in the dark and under simulated illumination using the two chemical potential method. The dark energy barriers reported here are those obtained using the same path tolerance and no climbing image, to match the parameters used for the illuminated runs.}
    \label{tab:dark-light}
\end{table}

We followed these NEB calculations with our same PDOS and L{\"o}wdin analysis; however, due to computational resources and convergence issues this was done only for the intrinsic cell. Table \ref{tab:qeff_TC} shows the results of  our $q_{eff}$ calculation, which revealed that in the presence of photo-excited carriers, the TSS\footnote{We were unable to reach convergence for our climbing image NEB calculations with the two chemical potential method. Therefore, this is not necessarily the true transition state structure. This is the highest energy image for our non climbing image runs} gets both of the enhancements we observed in the doped cells: First, the Si-DB is only singly occupied similar to the p-type cell in the dark. Second, the excited carrier occupies the BCH state which helps passivate the H as in the n-type cell in the dark. Combining the two effects reduces the Coulomb energy even further, and thus provides an even greater barrier reduction.  

\begin{table}[h]
    \centering
    \begin{tabular}{|c|c|c|c|c|c|}
    \hline
          $q_{H}$ & $q_{DB}$  & $q_{BC_a}$ & $q_{BC_b}$ & $q_{BCH}$ & $\Delta E_C$ \\
        \hline
          0.045 & -0.047 & 0.118 & 0.080 & 0.248 & 0.0031\\
         \hline
         
    \end{tabular}
    \vspace{5pt}
    \caption{Under illumination $q_{DB}$ is close to neutral as in the p-type dark calculation, and $q_H$ is also close to neutral as in the n-type dark calculation. Under illumination both effects are simultaneously achieved providing an even greater reduction to the barrier.}
    \label{tab:qeff_TC}
\end{table}

Here we wish to clarify that, in contrast to our earlier calculations, we were unable to obtain convergence with the lower path tolerance or with a climbing image. 
The TSS used for this calculation is not necessarily the true transition state structure; it is instead, the highest energy image from our non-climbing image NEB.
Additionally, it has been show that photo-excited carriers affect the phonon spectrum in Si \cite{twoChem}. Thus the ZPE correction obtained for our intrinsic cell in the dark is likely not valid. 
Despite these issues, our results show close correspondence with experimental results \cite{Santos-Johnson-Street-1991} and thus reveal the promise of combining the two chemical potential method with NEB to simulate dynamical processes in illuminated systems.

\section{Conclusions}
In this paper we reported on a comprehensive first principals computational study on the effect of the Fermi level and photo-excited carriers on defect generation via Si-H bond breaking in c-Si and a-Si:H. We used density functional theory to study the electronic properties of the atoms locally involved in defect generation, in p-type, intrinsic, and n-type cells both for dark and illuminated conditions. 

First, we determined the lowest energy interstitial configuration for hydrogen in our three cells. We found that near a Si dangling bond, the Si-Si bond center was lowest energy configuration even in n-type Si.
We found that in these bond center configurations the hydrogen was neutral and the charge delocalized to the neighboring Si. 
We used NEB to compute Si-H bond breaking barriers at different Fermi levels. We found an asymmetric reduction in the barriers that aligns closely with experimentally reported values. 
We determined that the change in the local Coulomb energy provided a consistent explanation for the barrier reduction in our doped cells.
We performed our Si-H bond breaking analysis in a-Si:H. We found that the distribution of defect energy levels resulted in a distribution of barriers reductions. We computed the effective Si-H bond breaking barrier as a function of Fermi level that matched well to experimental results \cite{BEYER2003}.
Finally, we used the two chemical potential method \cite{twoChem} to simulate the effect of illumination. We found a close correspondence to reported experimental observations, providing support to the method of combining the two chemical potential method with NEB to simulate dynamical process in illuminated materials.

\bibliographystyle{unsrt}
\bibliography{main}
\end{document}


\label{supp_mat}
\begin{center}
    \Huge{Supplementary Material}
\end{center}

For our Fermi level dependent Si-H bond breaking calculations we performed several additional calculations to confirm our methods and results. Here we present these calculations. In tables the use of (-) indicates a calculation for which convergence was not obtained.
\section{Interstitial H energy}
\begin{figure}[h]
     \centering
    \begin{subfigure}[t]{0.45\textwidth}
    \caption{}
    \vspace{0pt}
    \includegraphics[width = \textwidth]{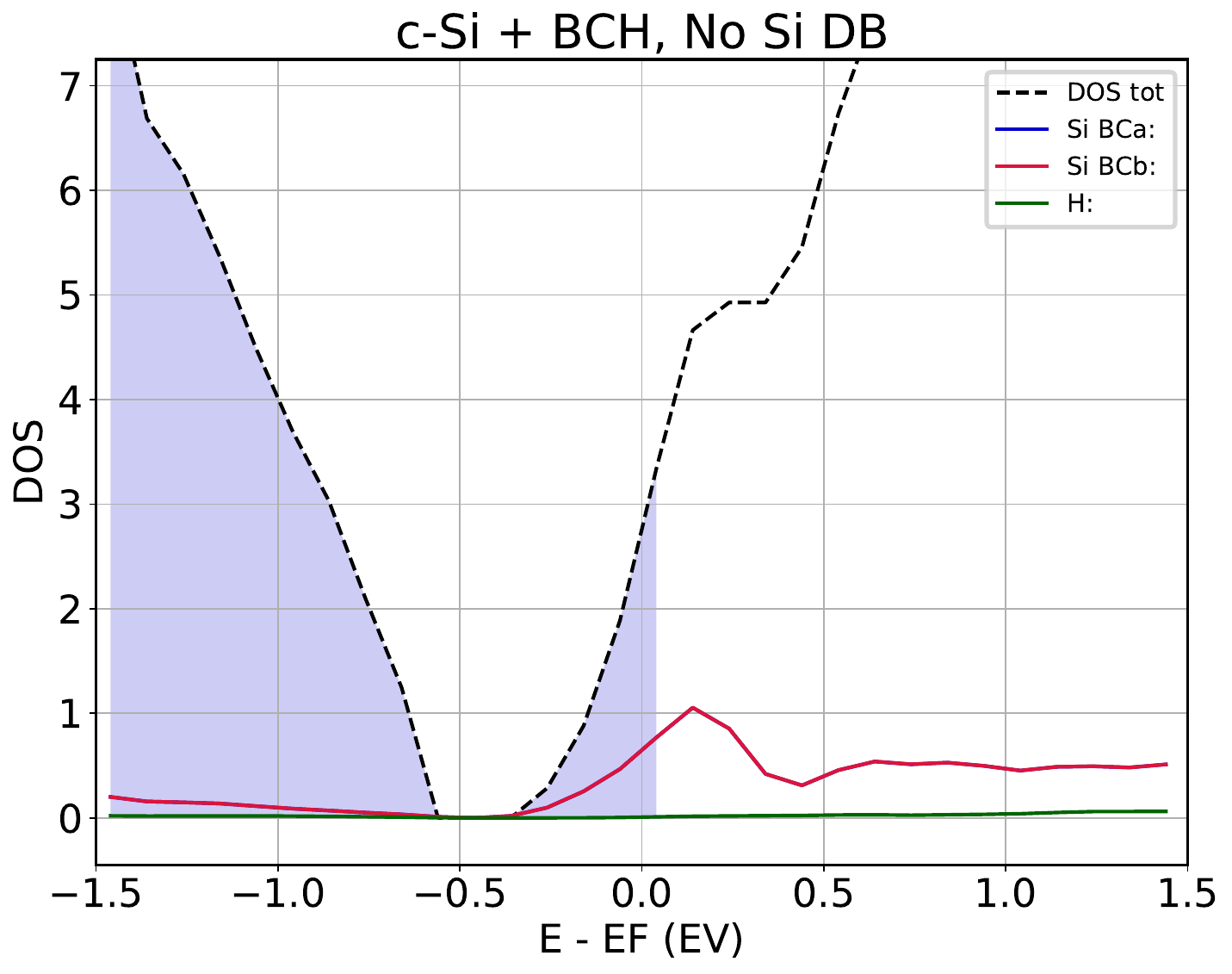}
    \label{fig:BCH_NoH}
    \end{subfigure}
    \hfill
     \begin{subfigure}[t]{0.45\textwidth}
    \caption{}
    \vspace{0pt}
    \includegraphics[width =\textwidth]{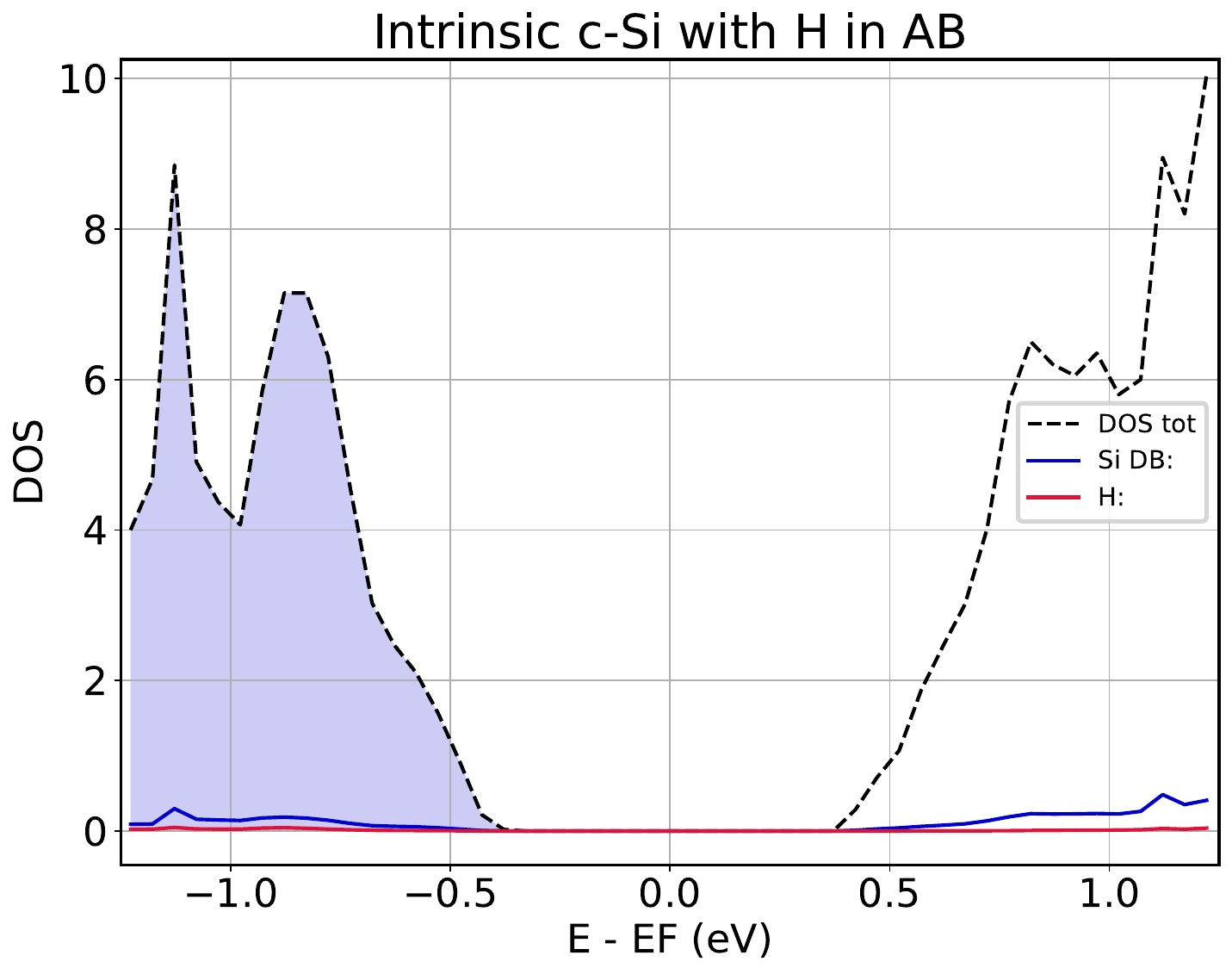}
    \label{fig:AB-PDOS}
    \end{subfigure}
    \begin{subfigure}[t]{0.45\textwidth}
    \caption{}
    \vspace{0pt}
    \includegraphics[width =\textwidth]{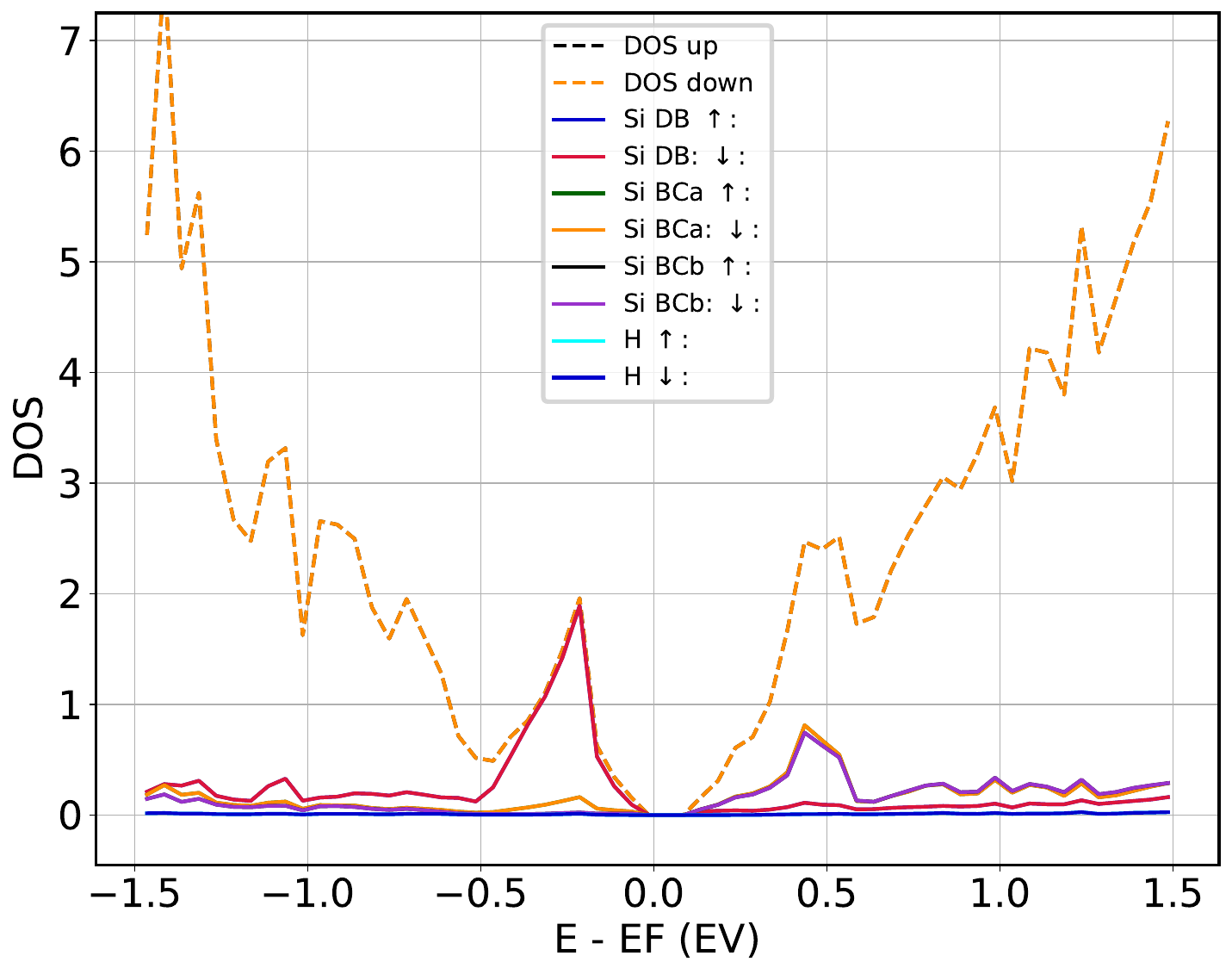}
    \label{fig:spin}
    \end{subfigure}
    \hfill
    \begin{subfigure}[t]{0.45\textwidth}
    \caption{}
    \vspace{0pt}
    \includegraphics[width =\textwidth]{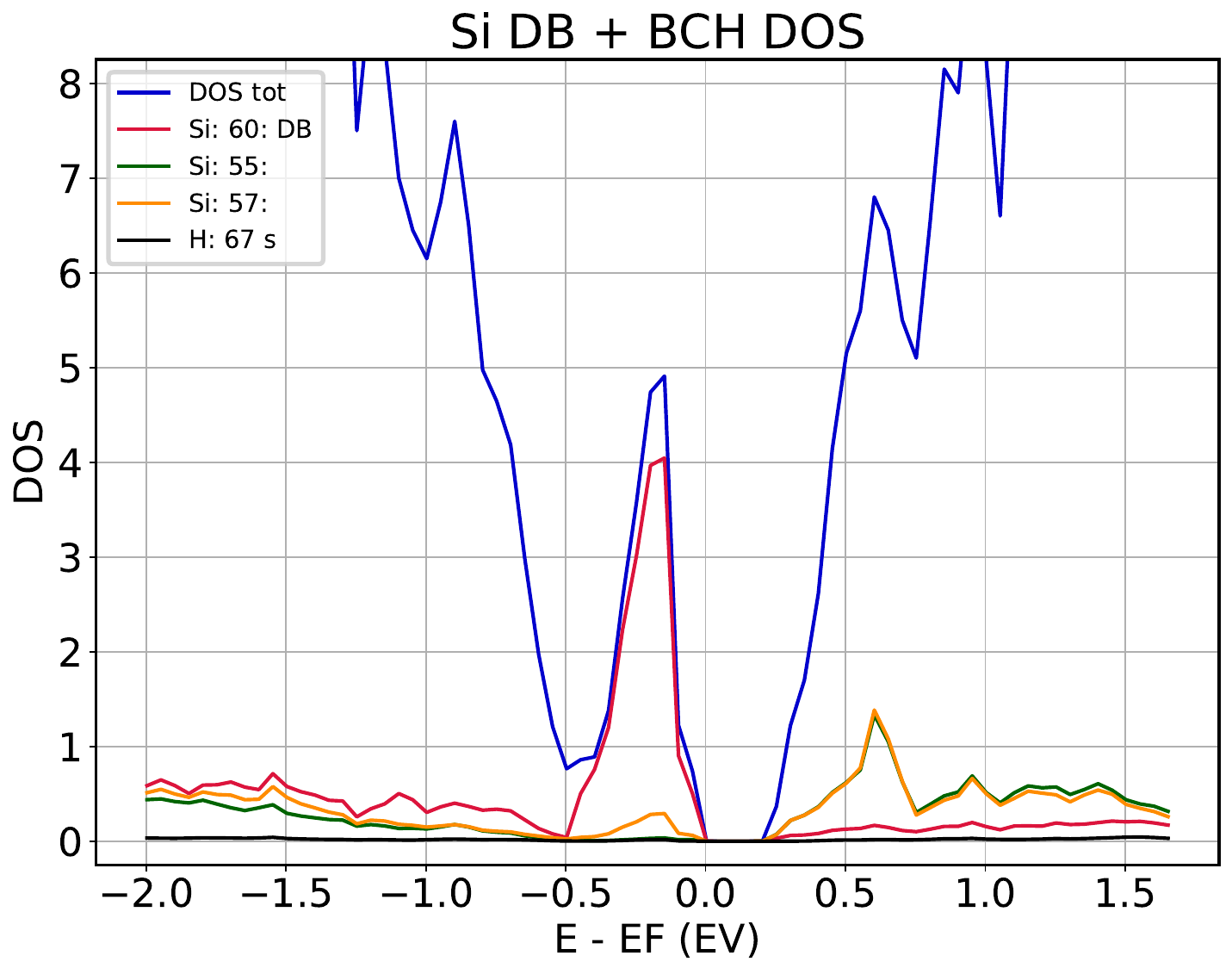}
    \label{fig:BCH_U}
    \end{subfigure}
    \vspace{-10pt}
    \caption{\SubCap{a} PDOS of perfet c-Si with a BCH complex. The state is in the exact same place but it is now occupied due to the lack of a Si DB at lower energy. \SubCap{b} PDOS for c-Si with a Si DB and the H at the AB interstitial site. The AB configuration is meta-stable and does not produce a mid gap state. \SubCap{c} Si DB + BCH final state analyzed using LSDA. This plot shows that the both spin states of the Si DB  are below the Fermi level, and thus the Si DB is doubly occupied. \SubCap{d} Even with U and U+V corrections, the Si DB remains doubly occupied. The BCH state is consistently $\sim 0.5$ eV higher than the Si DB.}
    \label{fig:cSi-multi-PDOS}
\end{figure}

Table \ref{tab:interH} show the energy for the three hydrogen interstitial locations: Si-Si bond center, tetrahedral, and anti-bonding, for all three super cell charges with a Si DB in the cell. The AB is the lowest energy for all super cell charges; however, this configuration is only meta-stable and does not result in any recombination active defect states, see Fig.~\ref{fig:AB-PDOS}. Additionally, the barrier for the H to return to the initial configuration is $\sim 1.1$ eV, which when compared to the $\sim 2.0$ eV barrier to migrate to a neighboring BC, makes this path very unlikely. The lowest energy defect creating configuration is the Si-Si BC for all cell types when a Si DB is present. Table \ref{tab:inter2} shows the energy of the interstitial H for the $Q=0$ and $Q=-q_e$ super cells in perfect c-Si. Similar to the work of Herring and Van de Walle \cite{Herring2001} we find that the tetrahedral is lowest in energy for the $Q=-q_e$ super cell. 
\begin{table}[h]
    \centering
    \begin{tabular}{|c|c|c|c|}
    \hline
         State & $Q=+q_e$ & $Q=0$ &  $Q=-q_e$  \\
         \hline
         BCH & -535.066558 Ry & -534.61178832 Ry & -534.07363748 Ry \\
         \hline
         Tetrahedral & - & -534.536213 Ry & -534.04636 Ry  \\
         \hline
         Anti-bonding & -535.163684 Ry & -534.68886373 Ry& -534.1500233 Ry \\
         \hline
         
    \end{tabular}
\vspace{20pt}
    \caption{Relaxed energies for all interstitial H configurations in c-Si with a DB, for $Q= +q_e$, $Q=0$, and $Q=-q_e$ super cells. Using a 3x3x3 k point grid, Ecutwfc = 50.}
    \label{tab:interH}
\end{table}

\begin{table}[h]
    \centering
    \begin{tabular}{|c|c|c|}
    \hline
         State & $Q=0$ &  $Q=-q_e$ \\
        \hline
         Initial & -718.86797 Ry & -718.3902 Ry \\
         \hline
         BC1 & -718.738226 Ry & -718.2659 Ry  \\
         \hline
         BC2 & -718.744297 Ry& -718.26927 Ry  \\
         \hline
         Tetrahedral & -718.70922 Ry & -718.25784 Ry  \\
         \hline
         Anti-bonding & -718.8283 Ry & -718.34833 Ry  \\
         \hline
         
    \end{tabular}
\vspace{20pt}
    \caption{Relaxed energies for all interstitial H configurations in perfect c-Si, for the Q = $0$ and Q = -$q_e$ super cells. Using a 3x3x3 k point grid, Ecutwfc = 50.}
    \label{tab:inter2}
\end{table}

\clearpage

\section{Si-H bond breaking c-Si}
Tables for the Si-H bond breaking barrier in c-Si for both P/B doped and charged cells.

\begin{table}[h]
    \centering
    \begin{tabular}{|c|c|c|}
    \hline
         Charge & $\Delta E_{forward}$ (eV) & $\Delta E_{reverse}$ (eV) \\
        \hline
        B doped & 1.77631 & 0.28931 \\
         \hline
         +1 & 1.79125 & 0.2903\\
         \hline
         0 & 2.0879 & 0.3057  \\
         \hline
         -1 & 2.0056 & 0.3146\\
         \hline
        P doped & 2.0104 & 0.3068 \\
         \hline
         
    \end{tabular}
\vspace{10pt}
    \caption{Final results for Si-H bond breaking barrier in c-Si comparing doped cell with charged cells. NEB parameters: path tol = 0.1 c-Si with PBESOL EXX and rrkjus PP, with CI.}
\end{table}

\begin{table}[h]
    \centering
    \begin{tabular}{|c|c|c|}
    \hline
         Charge & $\Delta E_{forward}$ (eV) & $\Delta E_{reverse}$ (eV) \\
        \hline
        +1 & 1.732 & 0.254\\
        \hline
        0 & 2.024 & 0.259\\
        \hline
        -1 & 1.925 & 0.268\\
        \hline
         
    \end{tabular}
\vspace{10pt}
    \caption{Final results for c-Si, NEB parameters: path tol = 0.1, PBESOL EXX and PAW PP, no CI.}
\end{table}

\begin{table}[h]
    \centering
    \begin{tabular}{|c|c|c|}
    \hline
         Charge & $\Delta E_{forward}$ (eV) & $\Delta E_{reverse}$ (eV) \\
        \hline
        +1 & 1.733 & 0.255\\
        \hline
        0 & 2.073 & 0.308\\
        \hline
        -1 & 1.986 & 0.346\\
        \hline
         
    \end{tabular}
\vspace{10pt}
    \caption{Final Si-H bond breaking results for c-Si. NEB parameters: path tol = 0.7 with PBESOL EXX and PAW PP, with CI.}
    \label{tab:Final}
\end{table}

\begin{table}[h]
    \centering
    \begin{tabular}{|c|c|c|c|}
    \hline
        Charge & $\Delta E_{forward}$ (eV) & $\Delta E_{reverse}$ (eV) & $\frac{E_f}{E_r}$ \\
        \hline
         +1 & 0.268 & 0.364 & 0.74  \\
         \hline
         0 & 0.293 & 0.375 & 0.78 \\
         \hline
         -1 & 0.279 & 0.307 & 0.91 \\
         \hline
    \end{tabular}
\vspace{10pt}
    \caption{Final BC-BC hopping results for c-Si. NEB parameters: path tol = 0.7 with PBESOL EXX and PAW PP, with CI.}
    \label{tab:bc_bc}
\end{table}
\vspace{50pt}
\subsection{Zero point energy correction}
The reported values in table \ref{tab:P1-BB-CH} were much higher than the expected values from our previous work, which showed an Si-H bond breaking barrier distribution centered around 1.3 eV. This distribution was calculated with MD in LAMMPS using SiH GAP potential. 
We explored various parameters to see what is the cause of the significant difference.

\begin{table}[h]
    \begin{tabular}{|c|c|c|c|c|c|}
    \hline
         Method & PP/Exc & Opt  & Path tol &  $\Delta E_{forward}$ (eV) & $\Delta E_{reverse}$ \\
        \hline
         DFT & NC/PBE & BFGS &  0.1  &  2.038154  & 0.381815  \\
         \hline
         DFT & NC/PBE & BFGS &  0.05 &   2.152296  & 0.495956  \\
         \hline
         DFT & NC/SLA & BFGS &   0.1 &   1.912620  & 0.342148  \\
         \hline
         DFT & NC/SLA & BFGS &  0.1 &   1.930685  & 0.360214  \\
         \hline
         MD & GAP & quick min &  $10^{-8}$ &   1.33974  & 0.238787  \\
         \hline
         MD & GAP & quick min &  $10^{-4}$ & 1.63737  & 0.51272  \\
         \hline
         
    \end{tabular}
\vspace{20pt}
    \caption{$2.04 g/cm^3$ a-Si:H Bond breaking NEB for path 1}
    \label{tab:gap-neb}
    
\end{table}
One important issue is the zero point energy correction ($\Delta ZPE$) to the thermal partition function. In the MD runs a damped dynamics minimization is used that allows for the NEB to be computed at non-zero temperature. For the DFT, the barriers are computed at the ground state; as such, $\Delta ZPE$ becomes an important contribution to the $\mathcal{Z}^{\ddag}/\mathcal{Z}^{A}$ term in Eq.~\ref{eqn:tst_1}. $\Delta ZPE$ is commonly computed using using the first order approximation
\begin{equation}
    \Delta ZPE = \sum_n \frac{1}{2} \hbar \omega_n^{\ddag} - \sum_m \frac{1}{2} \hbar \omega_m^A
\end{equation}
where $\omega^{\ddag}/\omega^A$ are the phonon modes at the transition state and initial state respectively. We computed the phonon spectrum for the initial, transition, and final state using \textit{phonon} package in QE. Due to the substantial computational expense, this was done for the intrinsic cell only. 
\begin{table}[h]
    \centering
    \begin{tabular}{|c|c|}
    \hline
            Barrier & $\Delta ZPE$ \\
            \hline
         $E_{forward}$ & -0.632 eV\\
         \hline
         $E_{reverse}$ & 0.053 eV\\
        \hline
         
    \end{tabular}
\vspace{10pt}
    \caption{$\Delta ZPE$ for Si-H bond breaking in intrinsic c-Si.}
    \label{tab:Supp-ZPE}
\end{table}

\section{Si-H Bond Breaking in a-Si:H} 
After determining the converged parameters were proceeded to compute the Si-H bond breaking barrier and BC-BC hopping barrier in several a-Si:H cells with varying mass density. We again used both B/P doped and charged cells.
\begin{table}[h]
    \centering
    \begin{tabular}{|c|c|c|c|}
    \hline
         Charge/Doping & $\Delta E_{forward}$ (eV) & $\Delta E_{reverse}$ & $E_F$\\
        \hline
         (i-type) Q=0 & 2.03815  & 0.381815 & 5.4157  \\
         \hline
         (n-type) P Doped & 2.028300 & 0.389923 & 5.6806 \\
         \hline
          (i-type) Q=-1 & 2.031770 & 0.41056 & 5.7991  \\
         \hline
         (p-type) B Doped & 1.974865 & 0.325204 & 4.8233\\
         \hline
         (i-type) Q=+1 & 2.014568  & 0.332619 & 4.870 \\
         \hline
         
    \end{tabular}
\vspace{10pt}
    \caption{ $2.04 g/cm^3$ a-Si:H Bond breaking NEB for path 1, No CI.}
    \label{tab:P1-BB-CH}
    
\end{table}

\begin{table}[h]
    \centering
    \begin{tabular}{|c|c|c|c|c|}
    \hline
         Charge/Doping & E Final (Ry) & $\mathrm{E_F}$ (eV)& HOMO (eV)& LUMO (eV) \\
        \hline
         (i-type) Q=0 & -457.1622 Ry & 5.0172 & 4.5610 & 5.2700\\
         \hline
         (n-type) P Doped& -462.5518 Ry & 4.9339 & 4.9287 & 4.9339\\
         \hline
         (p-type) B Doped & -454.5006 Ry & 4.6862 & 4.6740 & 4.7203\\
         \hline
         (i-type) Q=+1 & -457.5061 Ry & 4.7405 & 4.7365 & 4.7775\\
         \hline
         (i-type) Q=-1 & -456.7921 Ry & 4.8956 & 4.8954 & 4.9004\\
         \hline
         
    \end{tabular}
\vspace{10pt}
    \caption{$2.00 g/cm^3$ a-Si:H initial structure relaxation.}
    \label{tab:aSi200}
    
\end{table}

\begin{table}[h]
    \centering
    \begin{tabular}{|c|c|c|}
    \hline
         Charge/Doping & $\Delta E_{forward}$ (eV) & $\Delta E_{reverse}$ \\
        \hline
         (i-type) Q=0 & 2.14533 & 0.615198 \\
         \hline
         (n-type) P Doped & 2.2219 & 0.6465 \\
         \hline
          (i-type) Q=-1 & 2.204095 & 0.640397 \\
         \hline
         (p-type) B Doped & 2.177966 & 0.541463 \\
         \hline
         (i-type) Q=+1 & 2.189911  & 0.537928 \\
         \hline
         
    \end{tabular}
\vspace{10pt}
    \caption{ $2.00 g/cm^3$ a-Si:H Bond breaking NEB, No CI.}
    \label{tab:aSi200_2}
    
\end{table}

\begin{table}[h]
    \centering
    \begin{tabular}{|c|c|c|}
    \hline
         Charge/Doping & $\Delta E_{forward}$ (eV) & $\Delta E_{reverse}$ \\
        \hline
         (i-type) Q=0 & - & - \\
         \hline
         (n-type) P Doped & 0.64235 & 0.878876 \\
         \hline
         (p-type) B Doped & 0.619591 & 1.083476  \\
         \hline
         (i-type) Q=+1 & 0.639572  & 1.062325 \\
         \hline
         (i-type) Q=-1 &  & \\
         \hline
         
    \end{tabular}
\vspace{10pt}
    \caption{$2.00 g/cm^3$ a-Si:H BC-BC Hop NEB , No CI.}
    \label{tab:aSibcbc}
    
\end{table}

\begin{table}[h]
    \centering
    \begin{tabular}{|c|c|c|c|}
    \hline
        End & Charge & $\Delta E_{forward}$ (eV) & $\Delta E_{reverse}$ (eV) \\
        \hline
         BC1  & 0 & 2.320 & 0.680 \\
         \hline
         BC1  & -1 & 2.332 & 0.651  \\
         \hline
         BC1  & +1 & - & -  \\
         \hline
         BC1  & +2 & 2.042 & 0.549  \\
         \hline
         BC1  & BNQ & 2.270 & 0.653   \\
         \hline
         \hline
         BC2  & 0 & 2.184 & 0.909   \\
         \hline
         BC2  & -1 & 2.038 & 0.949   \\
         \hline
         BC2  & +1 & 2.278 & 0.846   \\
         \hline
         BC2  & +2 & 2.093 & 0.813   \\
         \hline
         BC2  & BNQ & 2.141 & 0.920  \\
         \hline
         \hline
         BC3  & 0 & 2.093 & 0.479  \\
         \hline
         BC3  & -1 & 2.060 & 1.185  \\
         \hline
         BC3  & +1 & 2.027 & 0.411  \\
         \hline
         BC3  & +2 & 1.780 & 0.362  \\
         \hline
         BC3  & BNQ & 2.054 & 0.466  \\
         \hline
         \hline
         BC4  & 0 & 2.074 & 0.807   \\
         \hline
         BC4  & -1 & 2.176 &  1.078   \\
         \hline
         BC4  & +1 & 2.252 & 0.970  \\
         \hline
         BC4  & +2 & 1.840 & 1.197  \\
         \hline
         BC4  & BNQ & 2.129 & 0.891  \\
         \hline

    \end{tabular}
\vspace{10pt}
    \caption{Si-H NEB barriers for 2.18 $g/cm^3$ a-Si:H, H44 using DFT+U with PBE Exx and NC PP}
    \label{tab:aSi218}
\end{table}

\begin{table}[h]
    \centering
    \begin{tabular}{|c|c|c|c|}
    \hline
        End & Charge & $\Delta E_{forward}$ (eV) & $\Delta E_{reverse}$ (eV) \\
        \hline
         BC1  & 0 & 2.139 & 0.625   \\
         \hline
         BC1  & $+q_e$ & 1.916 & 0.553   \\
         \hline
         BC1 & $-q_e$ & 2.026 & 0.642   \\
         \hline
         \hline
         BC2  & 0 & 2.047 & 0.720   \\
         \hline
         BC2  & $+q_e$ & 1.830 & 0.624   \\
         \hline
         BC2  & $-q_e$ & 2.034 & 0.706   \\
         \hline
         \hline
         BC3  & 0 & 1.790 & 0.089  \\
         \hline
         BC3  & $+q_e$ & 1.522 & 0.064   \\
         \hline
         BC3  & $-q_e$ & 1.736 & 0.139   \\
         \hline
         \hline
         BC4  & 0 & 1.697 & 0.646  \\
         \hline
         BC4  & $+q_e$ & 1.464 & 0.693   \\
         \hline
         BC4  & $-q_e$ & 1.642 & 0.908   \\
         \hline

    \end{tabular}
\vspace{20pt}
    \caption{2.21 $g/cm^3$ a-Si:H, Si-H NEB barriers for $Q = +q_e$, $Q = 0$, and $Q=-q_e$. Using PBEsol Exx with PAW PP.}
    \label{tab:221_neb}
\end{table}

\clearpage
\section{DFT+U/ DFT+U+V}
DFT+U+V uses the extended Hubbard model to add on-site and inter-site corrections to DFT. 
\begin{equation}
    H_{Hub} = \sum_{ij\sigma} t_{ij} c^{\dagger}_{i\sigma}c_{j\sigma} + U \sum_i n_{i \uparrow} n_{i \downarrow} \pm \sum_{ij \sigma \sigma'} V_{ij} n_{i \sigma} n_{j \sigma'}
\end{equation}

These corrections greatly improve localization.

\subsection{Computing U and U+V with RPA}
The first Step is to compute U and V. This is done using DFPT to compute the linear response matrix in the random phase approximation.
\begin{equation}
    \chi_{RPA} = \frac{\chi_0}{1 - U \chi_0}
\end{equation}

Which when inverted gives U on the diagonal and V on the off diagonal.
For 67 atom super cell there are 201 irreducible representations.
The wall time for HP calculation was 36h56m43.52s. 

Different H configurations using DFT + U with NC PP and PBE Exc, and using U Si-3p = 1.427, U H-1s = 6.36. Computed using hp.x

\begin{figure}
        \centering
     \begin{subfigure}[t]{0.45\textwidth}
    \caption{}
    \vspace{0pt}
    \includegraphics[width = \textwidth]{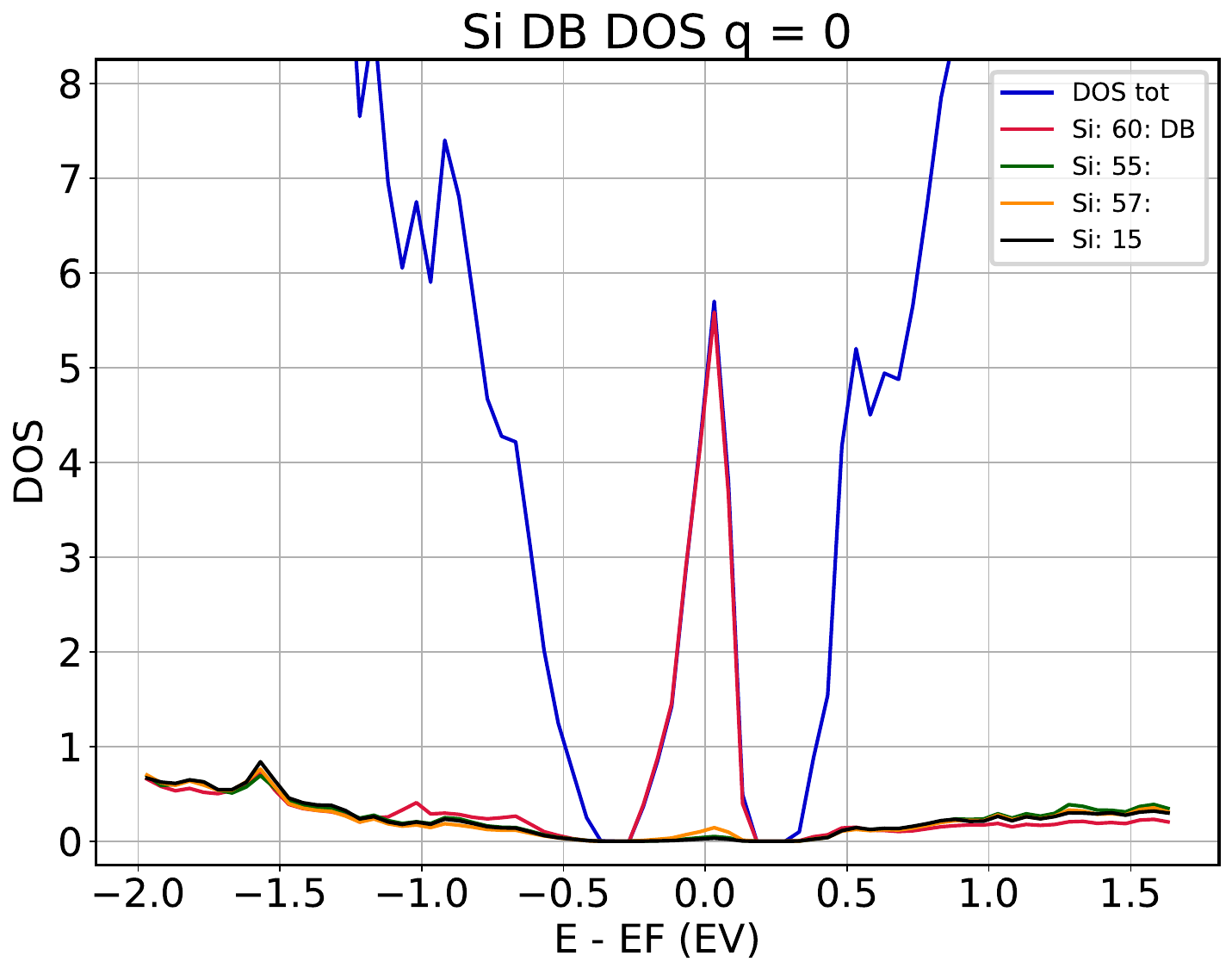}
    \label{fig:dbU}
    \end{subfigure}
    ~
    \begin{subfigure}[t]{0.45\textwidth}
    \caption{}
    \vspace{0pt}
    \includegraphics[width =\textwidth]{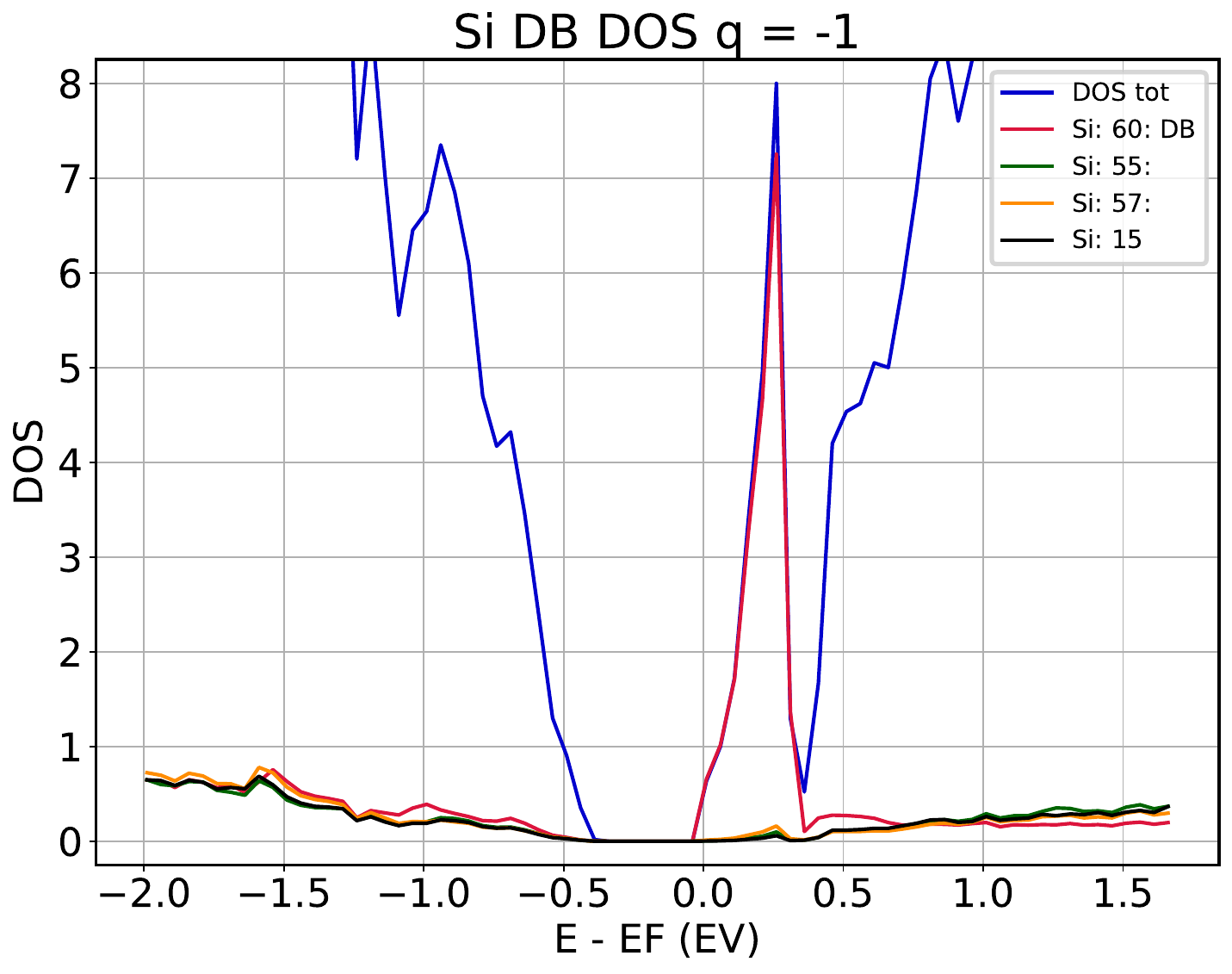}
    \label{fig:NQdbU}
    \end{subfigure}
    \caption{ Using the computed U the Si DB \SubCap{a} occupied by one electron is near mid gap  \SubCap{b} occupied by two electrons shifts the peak of the DB energy $\sim 0.2$ eV, which is consistent with reported values.}
    \label{fig:SiDB_U}
\end{figure}

Different H configurations using DFT + U + V with NC PP and PBE Exc, and using U Si = 1.427, U H = 6.360, V Si-Si = 0.210, V Si-H = 0.313 

\begin{table}[h]
    \centering
    \begin{tabular}{|c|c|c|c|c|}
    \hline
        H configuration & Total Charge  & $\mathrm{E_{tot}}$ & $\mathrm{E_{F}}$ & $\mathrm{E_{U}}$ \\
        \hline
          DB & 0  & -533.57177431 Ry  &  6.3661 eV & 2.82749 Ry  \\
         \hline
          DB & -1 & -533.06820708 Ry &  6.4336 eV & 2.883298 Ry\\
         \hline
         BCH & 0  & -534.61178832 Ry  & 6.3612 eV & 2.87437757 Ry  \\
         \hline
          BCH & -1 & -534.07363748 Ry & 6.6565 eV & 2.947886 Ry\\
         \hline
          Anti-bond & 0  & -534.68886373 Ry  & 6.3615 eV & 2.89003340 Ry  \\
         \hline
          Anti-bond & -1 & -534.1500233 Ry & 6.7380 eV & 2.95298 Ry\\
         \hline
         Tetrahedral & 0  & -534.536213 Ry  &  6.09336 eV &  2.91630647 Ry  \\
         \hline
          Tetrahedral  & -1 & -534.04636Ry & 6.2505 eV & 2.96175407 Ry\\
         \hline
         
    \end{tabular}
    \caption{BCH vs Tetrahedral H With q = 0 and q = -1 super cell using DFT+U+V.}
    \label{tab:UpV}
\end{table}

After gaining confidence with DFT+U+V, we proceeded to compute the Si-H bond breaking barriers. The use of Hubbard corrections slightly enhances the effect of the Fermi level on the barriers. Each configuration and atom type requires a full $\chi_{RPA}$ calculation which scales as $\mathrm{N^3ln(N)}$. The continued use of DFT+U+V was not realistic for this reason.
\begin{table}[h]
    \centering
    \begin{tabular}{|c|c|c|}
    \hline
         Charge & $\Delta E_{forward}$ (eV) & $\Delta E_{reverse}$ (eV) \\
         \hline
         +1 & 2.1539 & 0.4956\\
         \hline
         B doped & 2.1979 & 0.4989 \\
        \hline
         0 & 2.3840 & 0.4410  \\
         \hline
         -1 & 2.3062 & 0.5137\\
         \hline
         P doped & 2.3059 & 0.4724 \\
         \hline
         
    \end{tabular}
    \vspace{10pt}
    \caption{Results for c-Si with 3x3x3 super cell no CI with PBE-NC and +U. CI did not converge for these calculations}
\end{table}

\section{Supporting Figures}

\begin{figure}
        \centering
     \begin{subfigure}[t]{0.45\textwidth}
    \caption{}
    \vspace{10pt}
    \includegraphics[width = \textwidth]{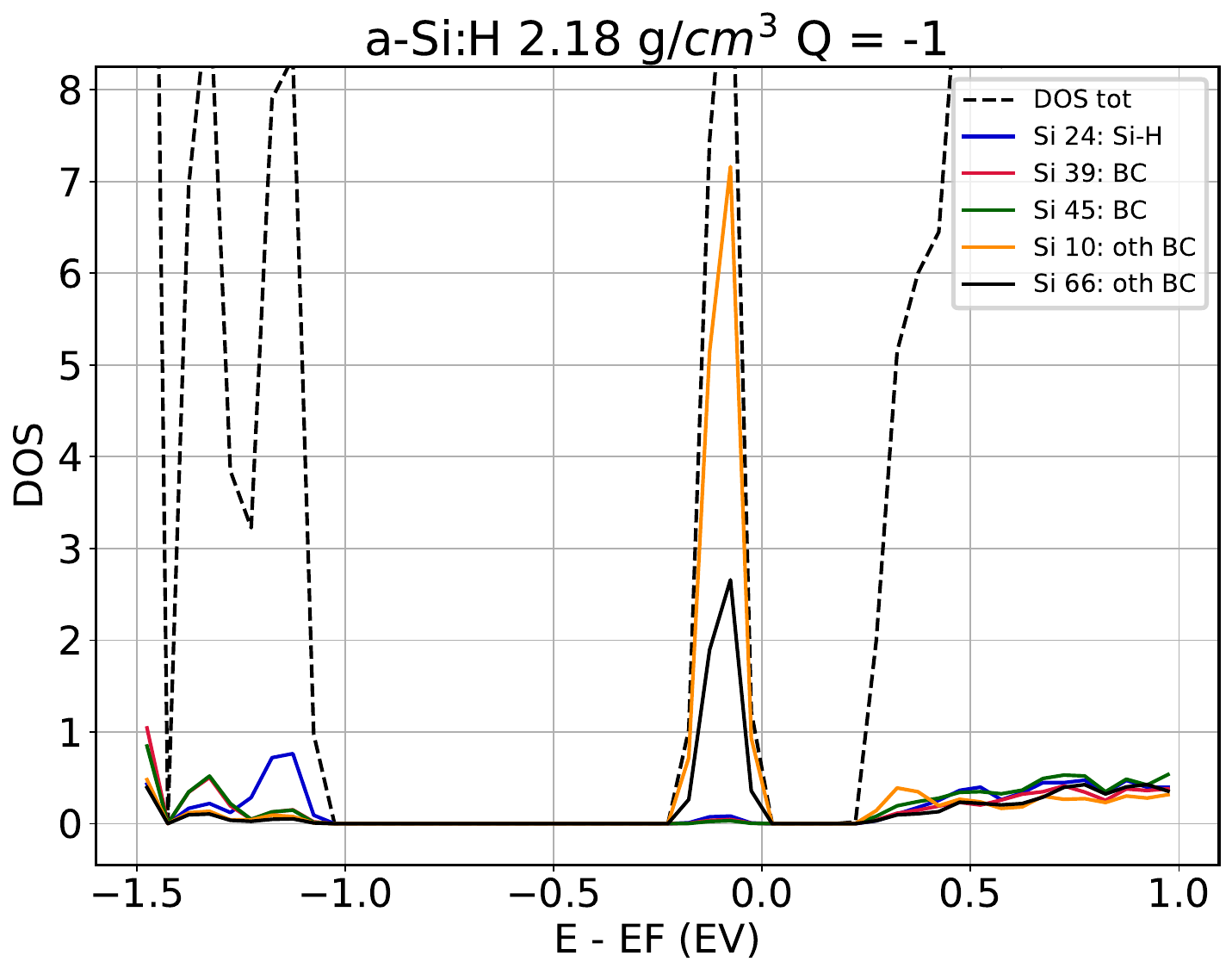}
    \label{fig:218_1}
    \end{subfigure}
    ~
    \begin{subfigure}[t]{0.45\textwidth}
    \caption{}
    \vspace{10pt}
    \includegraphics[width =\textwidth]{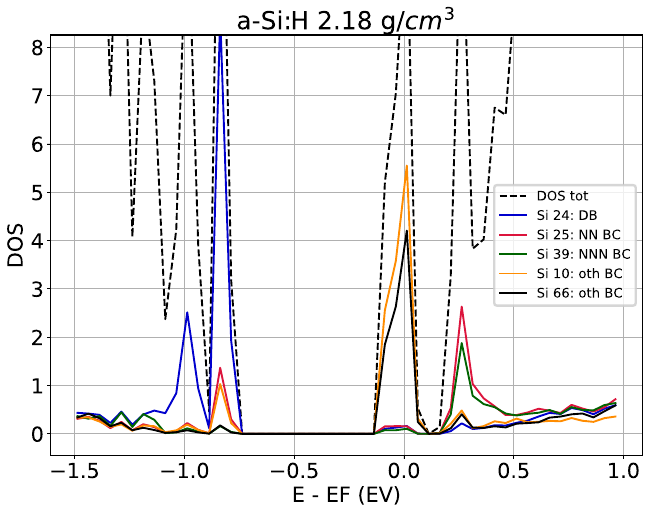}
    \label{fig:218_2}
    \end{subfigure}
    \hfill
    \begin{subfigure}[t]{0.45\textwidth}
    \caption{}
    \vspace{10pt}
    \includegraphics[width =\textwidth]{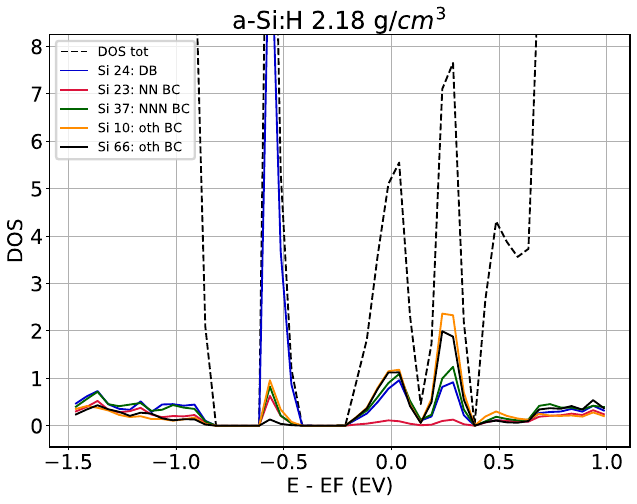}
    \label{fig:218_3}
    \end{subfigure}
    \hfill
    \begin{subfigure}[t]{0.45\textwidth}
    \caption{}
    \vspace{10pt}
    \includegraphics[width =\textwidth]{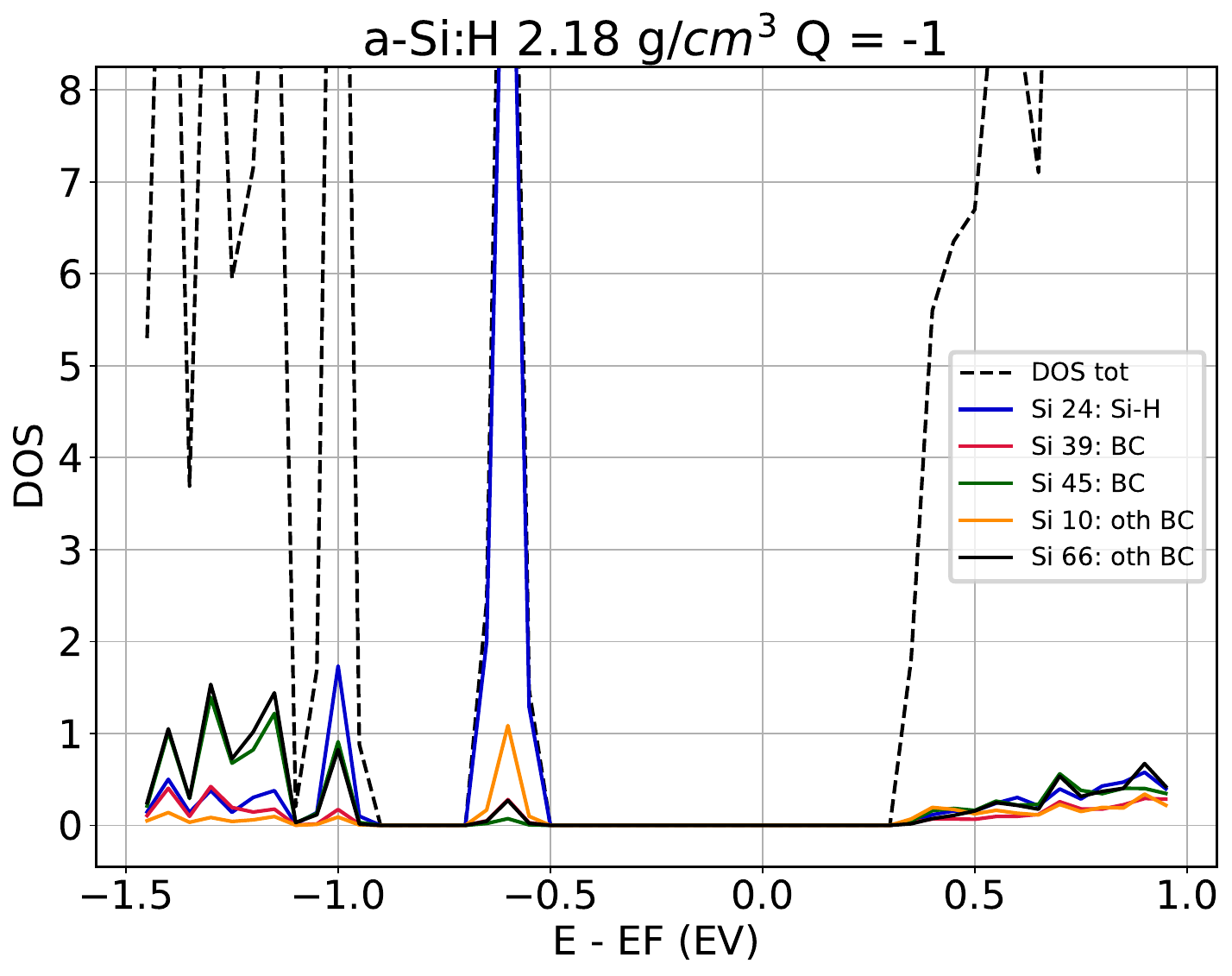}
    \label{fig:2184}
    \end{subfigure}
    \caption{ PDOS from one of the a-Si:H cells showing the variability in the energy of the Si DB state and the BCH state within the bandgap.}
    \label{fig:aSiH-multi-PDOS}
\end{figure}